\documentclass{aa}  
\usepackage{bm} 
\usepackage{mathrsfs}
\usepackage{graphicx}
\usepackage{txfonts}
\usepackage{xcolor}
\usepackage[colorlinks=true,citecolor=blue,linkcolor=blue,]{hyperref}
\usepackage{multirow}
\usepackage{mathrsfs}
\usepackage{lineno}
\usepackage{makecell}
\usepackage{afterpage}

\newcommand{\rorb}{r_\mathrm{orb}}
\newcommand{\ri}{r_\mathrm{i}}
\newcommand{\ro}{r_\mathrm{o}}

\begin{document}

\title{Tidal locking as a negative feedback on Earth-like planetary dynamos: consequences for magnetic shielding and habitability}
\titlerunning{Tidal locking as a negative feedback on Earth-like planetary dynamos}

\author{J. P. Hidalgo\inst{1,2}\thanks{\email{juanpablo.hidalgo@uniroma1.it}} \and D. R. G Schleicher\inst{1}}


\institute{Dipartimento di Fisica, Sapienza, Università di Roma, Piazza le Aldo Moro 5, 00185 Roma, Italy. \and 
INAF, Observatory of Abruzzo, Via Mentore Maggini snc, I-64100 Teramo, Italy}

   \date{Received XXX; accepted XXX}
 
  \abstract
  {We investigate how tidal locking affects the magnetic field generation and magnetospheric protection of Earth-like planets orbiting the habitable zone of M dwarfs. While the role of stellar activity has been widely studied, the impact of tidal locking on the planetary dynamo remains largely unexplored. We develop a framework that couples the stellar wind with two distinct dynamo paradigms: direct rotational scaling of a dipolar field, and energy-flux scaling where the local Rossby number governs the magnetic geometry, capturing a rapid transition from a dipolar to a multipolar configuration. We employ a Constant Time Lag model to capture the continuous tidal spin-down of the planets, evaluating two distinct planetary interiors: a highly dissipative Modern Earth and a less dissipative, rapidly rotating Early Earth. We find that tidal locking acts as a severe negative feedback on planetary magnetism across both paradigms, where the dipolar field frequently collapses before full tidal synchronization is reached. At the inner boundary, nearly all the planets are rapidly locked and the intrinsic magnetic fields are significantly attenuated. The outer boundary provides a favorable environment only for early M dwarfs ($M_\star \gtrsim 0.32~M_\odot$) hosting Early Earth. Here, the combination of non-synchronous rotation and residence outside the sub-Alfvénic regime provides atmospheric protection likelihoods of $97-100\%$. For mid-to-late M dwarfs, the combination of inevitable tidal locking and a sub-Alfvénic environment results in a total collapse of the magnetic shield, reducing the likelihood of atmospheric protection to essentially zero across the entire habitable zone.}


\maketitle
\nolinenumbers

\section{Introduction}

M dwarfs constitute the predominant stellar population in our galaxy, accounting for approximately 70–75\% of all stars \citep{Henry2006, Bochanski2010}. Given their abundance and very long main-sequence lifetimes, these low-mass stars are increasingly recognized as key targets in the search for potentially habitable exoplanets \citep{2022planetMdwarf}. The habitable zone (HZ), defined as the region where a terrestrial planet can maintain liquid water on its surface \citep{Kasting1993, Selsis2007}, is located significantly closer to the host star in M-dwarf systems compared to other stars. This proximity provides a good contrast to detect exoplanets via transit photometry \citep{Deeg2018}, and a larger radial velocity signal, within the current spectrograph sensitivities \citep{Kasting2014}.


Within these close-in habitable zones, terrestrial planets are highly susceptible to tidal locking \citep{Griessmeier2005, Leconte2015,Shields2016, Barnes2017, Childs2022}. Such a state has been considered to potentially make the question of habitability more severe, as it creates extreme thermal gradients between a constantly irradiated day-side and a frozen night-side. On the other hand, climate models have shown that atmospheric convection and efficient energy transport can strongly alleviate these extreme temperatures. Furthermore, stabilizing feedback from cloud formation can even increase the size of the habitable zone and move its inner region closer to the star \citep{Yang2013, Yang2014,2014PNAS..111..629H, Kopparapu2016}.


However, while the thermal stability of planets is a primary concern, the long-term retention of an atmosphere depends critically on the magnetic protection from stellar wind erosion \citep{See2014}. Within our Solar System, the atmospheric depletion of Mars serves as a primary example of the consequences of a failed global dynamo \citep{Lammer2008,Lammer2013, Jakosky2015, Dong2018}. Although several studies have explored the implications of stellar winds on the magnetospheres, the intrinsic planetary magnetic field is typically assumed to be a fixed input parameter \citep{Vidotto2013, Hidalgo2025b}. One important physical effect that is frequently overlooked is the fundamental dependence on rotation of the planetary dynamo. Some studies have explored the impact of tidal locking on the planetary magnetism using rotation-dependent scaling, which reduces the magnetic dipolar moments significantly \citep{Griessmeier2004,Griessmeier2005,2007AsBio...7..185L,Griessmeier2009, RodriguezMozos2019}. More complete models, such as that of \citet{Zuluaga2013}, have improved upon this by coupling thermal evolution with updated dynamo scaling laws, showing that slow rotation drives the magnetic field into a less protective multipolar regime. However, these studies often rely on simplified tidal-locking criteria or fixed rotation states, and they do not account for the gradual spin-down of the planet toward synchronous rotation.

 
In this work, we explore the link between tidal locking and planetary magnetism by developing a self-consistent framework that couples empirical stellar wind evolution with two distinct dynamo paradigms: direct rotational scaling and energy-flux scaling. This paper is organized as follows, in Section~\ref{model}, we detail our theoretical framework and the parameterization of our dynamo models. Our results for both a theoretical grid and a sample of observed stellar systems are presented in Section~\ref{results}. We discuss the limitations and the remaining uncertainties of our models in Section~\ref{discussions}, and our main conclusions are summarized in Section~\ref{conclusions}.

\section{The model \label{model}}
In this section, we describe the theoretical framework used to estimate the stellar HZ, the size of the upstream magnetosphere of Earth-like planets orbiting M dwarfs, the adopted scaling relations and assumptions on the planetary dynamo as well as the approach to model tidal locking. 

\subsection{Limits of the habitable zone}
To estimate the extension of the HZ, we used the 1D radiative-convective cloud-free climate model of \cite{Kopparapu2013,Kopparapu2014}, adopting the “runaway greenhouse” and the “maximum greenhouse” limits for the inner ($\ri$) and outer ($\ro$) boundaries, respectively. However, planetary rotation is a key parameter to determine the atmospheric circulation, and therefore, the mean planetary climate \citep[e.g.][]{Showman2013, Kaspi2015}. As tidally locked planets with low eccentricity are likely to be synchronous rotators, their rotation rates are typically significantly lower than Earth's. This can result in a weakened Coriolis force that drives strong convection at the substellar point, forming a thick, highly reflective permanent cloud shield, which moves the inner edge of the HZ significantly closer to the host star \citep{Yang2013, Yang2014, Way2015}. To quantify this effect, we adopted the estimated flux received by a slowly rotating planet predicted by the 3D models of \cite{Yang2014}, which yields a tidal locking inner boundary $r_\mathrm{i}^\mathrm{TL}$. We note that $r_\mathrm{i}^\mathrm{TL}$ represents a somewhat optimistic limit, as if self-consistent relations between metallicity, effective temperature, and the planetary rotational period are considered, then the limit of \cite{Yang2014} could overestimate the planetary albedo, pushing the true inner edge to lower stellar fluxes due to updated water-vapor absorption and the breakdown of the cloud shield at the shorter orbital periods required for close-in HZs \citep{Kopparapu2016}.

\subsection{Planetary magnetosphere}

The magnetosphere of a planet acts as an intrinsic shield, protecting the planetary atmosphere from the erosion produced by space weather \citep[e.g.][]{2007AsBio...7..185L,See2014}. The outer boundary of this region is known as magnetopause, and it corresponds to the distance from the planetary center where the stellar pressure $p_\star$ balances the planetary pressure $p_\mathrm{pl}$, defining the boundary between the planetary environment and the influence of the host star. A relevant contribution to the planetary side comes from the magnetic pressure $B_\mathrm{pl}^2/2\mu_0$, where $B_\mathrm{pl}$ is the magnetic field of the planet and $\mu_0$ is the magnetic permeability of vacuum. The pressure exerted by atmospheric escape should be considered too. In Earth's case this contribution can be safely neglected, as its atmospheric mass loss rate $\dot{M}_\mathrm{pl}$ is relatively low \citep[$\sim$ 1.4 kg/s;][]{Gunell2018, Hazra2025}. We note that mass loss can be significant in highly irradiated planets \citep{Ehrenreich2011, Owen2012, 2018PNAS..115..260D, 2019A&A...624L..10J}. However, the present model assumes that magnetic pressure dominates over atmospheric ram pressure in all cases, corresponding to the regime $\dot{M}_\mathrm{pl} \ll 10^5$ kg/s \citep[see][]{Hidalgo2025b}.

The stellar wind induces currents on the magnetosphere \citep{Chapman1931,Chapman1941,Ganushkina2018}. These currents generate a 
magnetic field $B_\mathrm{mc}$ opposite to the stellar magnetic field, increasing the shielding provided by the intrinsic planetary magnetic field. One way to quantify this effect is through the form factor $f_0$ \citep[e.g.][]{Griessmeier2004}, where $B_\mathrm{pl} + B_\mathrm{mc} \equiv 2 f_0 B_\mathrm{pl} $. The value of $f_0$ depends on the chosen geometry, e.g., $f_0 = 1$ ($f_0 = 1.5$) for a planar (spherical) magnetosphere, and $f_0 = 1.2$ for an ellipsoid \citep{Chapman1931,Russell2001}. Here, following \cite{volland1995handbook} we adopt $f_0 = 1.16$ for a “realistic” shape, closer to an ellipsoid due to the compression of the upstream side.

As the magnetic fields of exoplanets remain largely unconstrained \citep[][]{Brain2024, strugarek2025}, a dipolar configuration will be assumed, based on the large-scale magnetic field of Earth, i.e. \citep{Olson2006, Vidotto2013}
\begin{equation}
    B_\mathrm{pl}(r) = \frac{1}{2} B_\mathrm{p,0} \left( \frac{r_\mathrm{pl}}{r}\right)^3, \label{planetB}
\end{equation}
where $B_\mathrm{p,0}$ is the magnetic field at the pole, and $r_\mathrm{pl}$ is the planetary radius. With these considerations, the size of the upstream magnetosphere of a planet orbiting at a distance $r_\mathrm{orb}$ from the host star, is given by
\begin{equation}
    \frac{r_\mathrm{M}}{r_\mathrm{pl}}(\rorb) = \sqrt[6]{\frac{f^2_0 B_\mathrm{p,0}^2}{2\mu_0 p_\star (\rorb)}}.\label{rm}
\end{equation}

\subsection{Stellar pressure acting on the planet}
On the stellar side, the total pressure is given by $p_\star = p_\star^\mathrm{mag} + p_\star^\mathrm{ram} + p_\star^\mathrm{th}$, which includes contributions of the magnetic, ram, and thermal pressures, respectively. The stellar magnetic pressure is
\begin{equation}
    p_\star^\mathrm{mag}(\rorb) = \frac{B_\star(\rorb)^2}{2\mu_0} \label{pmag}
\end{equation}
where $B_\star$ is the stellar magnetic field. To model the topology of this quantity, we adopt a simplified Potential Field Source Surface approach \citep[e.g.][]{1969SoPh....9..131A, Jardine2002, Lang2012}. Within a reference radius $R_\mathrm{ref}$, the magnetic field is dominated by closed coronal loops and it decays as a dipole. Beyond $R_\mathrm{ref}$, the stellar wind stretches the magnetic field lines into a radial geometry, starting from a reference field given by $B_\mathrm{ref} = B_\star(R_\mathrm{ref}) = B_0 (R_\star/R_\mathrm{ref})^3 $, where $B_0$ is the average magnetic field at the stellar surface. Following \cite{Vidotto2013}, we set $R_\mathrm{ref} = 2.5 R_\star$, where $R_\star$ is the stellar radius, and therefore
\begin{equation}
        B_\star(r) = \left\{ \begin{array}{lcc} B_0 \left( \frac{R_\star}{r} \right)^3 & \text{if} & r \leq 2.5 R_\star \\ 0.4 B_0 \left( \frac{R_\star}{r}\right)^2  & \text{if} & r > 2.5 R_\star \end{array} \right. . \label{stellar-B}
\end{equation}

The stellar ram pressure is given by
\begin{equation}
    p_\star^\mathrm{ram}(r) = \rho (r) v_\mathrm{rel}^2(r) \label{pram}
\end{equation}
where $v_\mathrm{rel}$ is the velocity of the planet relative to the stellar
wind. Assuming isotropic conditions for the stellar wind, the density is
\begin{equation*}
    \rho(r) = \frac{\dot{M}_\star}{4\pi v_r(r) r^2}.
\end{equation*}
where $\dot{M}_\star$ is the stellar mass-loss rate and $v_r$ is the radial component of the velocity. These quantities are difficult to constrain in M dwarfs, mainly due to observational challenges involved in their detection \citep[e.g.][]{Wood2004,Jardine2019}.

In this model, for the mass-loss rate we use the relation based on the X-ray flux $\dot{M}_\star \propto F_\mathrm{X}^{0.77}$
reported by \cite{Wood2021}, which in terms of the X-ray luminosity ratio $R_\mathrm{X} = L_\mathrm{X}/L_\mathrm{bol}$ is given by
\begin{equation*}
    \dot{M}_\star = \dot{M}_\odot \left( \frac{R_\star}{R_\odot}\right)^{0.46} \left( \frac{R_\mathrm{X} L_\mathrm{bol}}{R_\mathrm{X,\odot} L_\mathrm{bol,\odot}}\right)^{0.77}
\end{equation*}
where $L_\mathrm{bol}$ is the bolometric luminosity, $\dot{M}_\odot = 2\cdot 10^{-14}~M_\odot\,\mathrm{yr}^{-1}$ and $R_\mathrm{X,\odot} = 10^{-6.24}$ \citep{2003ApJ...593..534J}. In late-type stars, the ratio $R_\mathrm{X}$ increases with rotation \citep{1981ApJ...248..279P} until it saturates at high rotation rates. This effect divides the active stars into two regimes: saturated and unsaturated \citep{2003A&A...397..147P, Wright2011, Wright2018}, where
\begin{equation*}
        R_\mathrm{X} = \left\{ \begin{array}{lcc} R_\mathrm{X, sat} & \text{if} & \mathrm{Ro} \leq \mathrm{Ro}_\mathrm{sat}^\mathrm{crit} \\ R_\mathrm{X,sat} \left(\mathrm{Ro}/\mathrm{Ro}_\mathrm{sat}^\mathrm{crit}\right)^\beta  & \text{if} & \mathrm{Ro} > \mathrm{Ro}_\mathrm{sat}^\mathrm{crit} \end{array} \right. ,
\end{equation*}
where $\mathrm{Ro} = P_\mathrm{rot}/\tau_\mathrm{conv}$ is the global Rossby number, $P_\mathrm{rot}$ is the rotation period of the star, $\tau_\mathrm{conv}$ is the convective turnover time, $R_\mathrm{X,sat}$ and $\mathrm{Ro}_\mathrm{sat}^\mathrm{crit}$ correspond to the saturated X-ray ratio, and the critical value of the Rossby number where the transition happens, respectively. For these last quantities, we use the results of \cite{Wright2011}, i.e. $R_\mathrm{X,sat}\approx 7.41 \cdot 10^{-4}$, $\mathrm{Ro}_\mathrm{sat}^\mathrm{crit} \approx 0.16$ and $\beta \approx -2.70$, which are values consistent with both, partially convective and fully convective M dwarfs \citep{Wright2018}. For the radial component of the velocity, we based on the $\beta$-velocity law of hot stellar winds \citep[e.g.][]{Castor1975, 2006MNRAS.370..580K}, and define
\begin{equation*}
    v_r(r) = v_\infty \left( 1 - \frac{R_\star}{r} \right)^\gamma,
\end{equation*}
where $v_\infty$ is the terminal velocity. For slow winds, a fair assumption is $v_\infty \approx v_\mathrm{esc}$ \citep{Cranmer2011, See2014, Suzuki2018, Hidalgo2025b}, where $v_\mathrm{esc} = \sqrt{2GM_\star/ R_\star}$ is the escape velocity. As M-dwarf winds are thought to be predominantly driven by Alfvén waves \citep{1980ApJ...242..260H, Vidotto2010}, the wind is rapidly accelerated near the star by efficient momentum transfer from the waves to the plasma \citep[e.g., see Fig. 1 of][]{Mesquita2020}. To capture this steep velocity gradient, we adopt an acceleration parameter of $\gamma = 1$, which ensures the wind quickly approaches its terminal velocity in the inner stellar system. The relative wind velocity is given by
\begin{equation*}
    \bm{v}_\mathrm{rel}(r) = v_r \bm{\hat{r}} + (v_\phi - v_\mathrm{orb})\bm{\hat{\phi}},
\end{equation*}
where $v_\phi$ is the azimuthal component of the velocity, and $v_\mathrm{orb} = \sqrt{GM_\star/r_\mathrm{orb}}$ is the orbital velocity, where $G$ is the gravitational constant. The azimuthal component becomes significant in planets with close-in orbits, where the interaction between the stellar wind and the planet is fundamentally non-radial. In the sub-Alfénic regime, i.e. $r < R_\mathrm{A}$ where $R_\mathrm{A}$ is the Alfvén radius, the magnetic field dominates the dynamics and the plasma roughly co-rotates with the star, leading to $v_\phi \approx \Omega_\star r$ \citep{Strugarek2018,RodriguezMozos2019}, where $\Omega_\star$ is the stellar rotation. Outside this radius, the kinetic energy of the plasma dominates the influence of the magnetic field, and the azimuthal component decays with $\propto 1/r$. With these considerations:
\begin{equation*}
     v_\phi(r) = \left\{ \begin{array}{lcc} \Omega_\star r & \text{if} & r \leq R_\mathrm{A} \\ \frac{\Omega_\star R_\mathrm{A}^2}{r}  & \text{if} & r > R_\mathrm{A} \end{array} \right. .
\end{equation*}

As the Alfvén radius is the distance where the magnetic energy density $B_\star^2/2\mu_0$ equals the kinetic energy density $1/2 \rho v_r^2$ \citep{Belenkaya2015}, it yields
\begin{equation*}
    R_\mathrm{A} \approx  \sqrt{\frac{4\pi B_\mathrm{ref}^2 R_\mathrm{ref}^4 }{\mu_0  \dot{M}_\star v_r}} \approx 0.4R_\star \sqrt{\frac{4\pi B_0^2 R_\star^2 }{\mu_0 \dot{M}_\star v_\mathrm{esc}}}.
\end{equation*}
The thermal pressure for a fully ionized hydrogen plasma is given by
\begin{equation}
    p_\star^\mathrm{th} = 2 \frac{\rho(r)}{m_\mathrm{p}} k_\mathrm{B} T(r), \label{pth}
\end{equation}
where $k_\mathrm{B}$ is the Boltzmann constant, $m_\mathrm{p}$ is the mass of the proton, and $T(r)$ is the stellar temperature. Following \cite{Johnstone2015b,Johnstone2015a}, the temperature profile is given by
\begin{equation}
    T(r) = (0.054 T_0 - 0.010) \left( \frac{\rho(r)}{\rho(1~\mathrm{AU})} \right)^{\alpha - 1},
\end{equation}
where $\alpha = 1.51$, and both the wind $T(r)$ and the base $T_0$ temperatures are expressed in megakelvin (MK). The base temperature is assumed to be proportional to that of the stellar corona, such that $T_0 \approx 0.75 T_\mathrm{cor}$. \cite{2015A&A...578A.129J} found that the coronal temperature can be expressed as
\begin{equation}
        \overline{T}_\mathrm{cor} = \left\{ \begin{array}{lcc} \overline{T}_\mathrm{cor,\odot} \left(  \frac{M_\star}{M_\odot} \right)^{-0.42} \left( \frac{\Omega_\star}{\Omega_\odot} \right)^{0.52} & \text{if} & \Omega_\star \leq \Omega_\mathrm{sat} \\ \overline{T}_\mathrm{cor,\odot} \left( \frac{M_\star}{M_\odot}\right)^{0.6} & \text{if} & \Omega_\star > \Omega_\mathrm{sat}  \end{array} \right. ,
\end{equation}
where $\Omega_\odot = 2.67\cdot 10^{-6}~\mathrm{rad\, s^{-1}}$, $\overline{T}_\mathrm{cor,\odot} \approx 2 ~\mathrm{MK}$, and
\begin{equation}
    \Omega_\mathrm{sat} = 13.53 \left( \frac{M_\star}{M_\odot} \right)^{1.08} \Omega_\odot \label{omega-sat}
\end{equation}
is the saturated rotation rate, obtained from the saturated Rossby number found by \cite{Wright2011} and the results of \cite{Reiners2014}.

\begin{figure*}[t!]
    \centering
    \includegraphics[width=0.9\hsize]{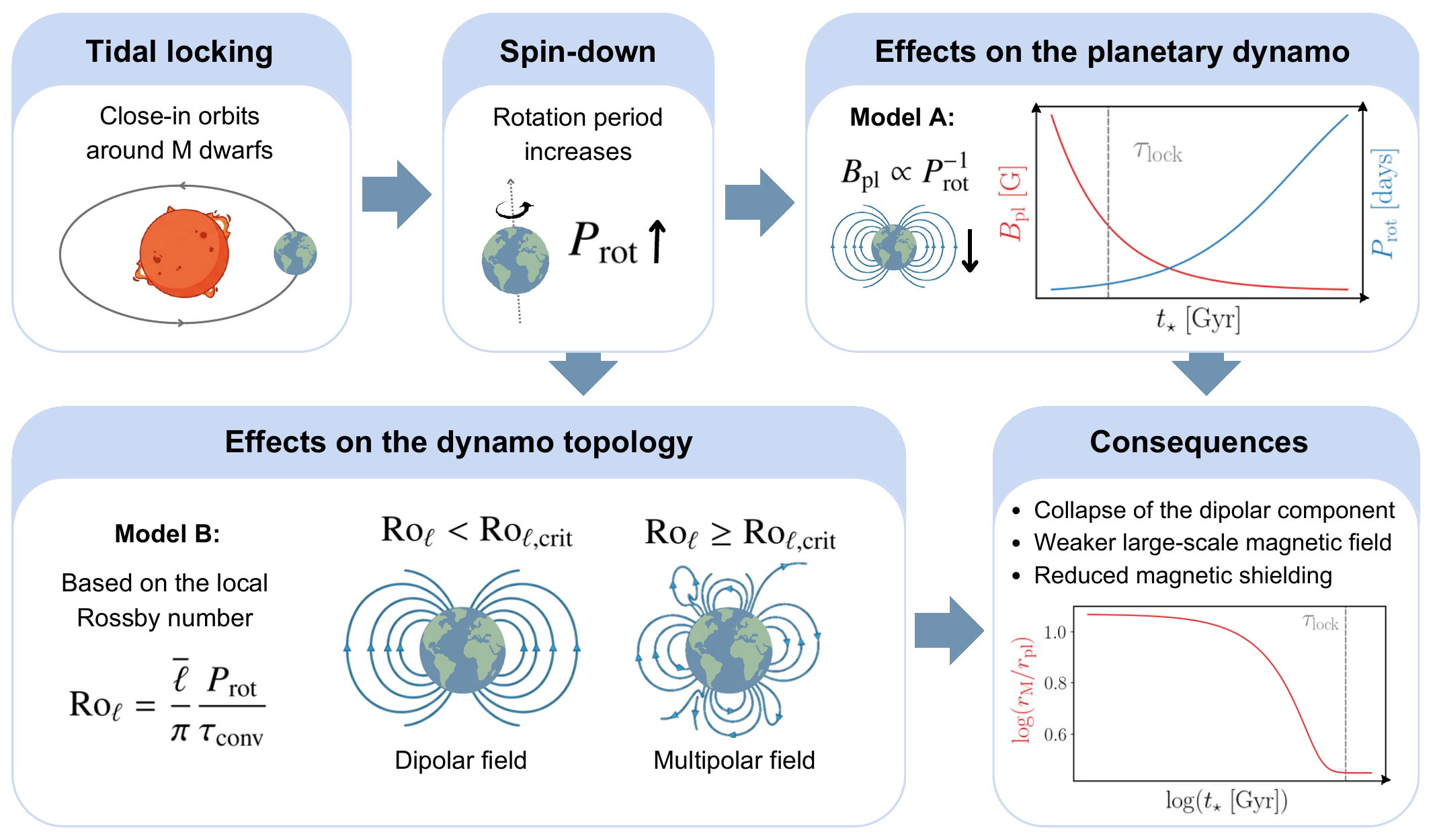}
    \caption{Summary of the main concepts presented here: Tidal locking influences the rotation period $P_\mathrm{rot}$ of the planet. The resulting spin-down directly influences the planetary magnetic field strength $B_\mathrm{pl}$ in Model A, while in Model B it affects the local Rossby number $\mathrm{Ro}_\ell$ (see Section~\ref{modelB}), which governs the geometry of the dynamo. Weak dipolar components decrease the magnetic protection of the planet.}
    \label{fig:illu}
\end{figure*}

\subsection{Planetary rotational evolution}
To ensure that the planetary rotation smoothly approaches synchronization, we solve the angular evolution equation from the classic Constant Time Lag (CTL) models \citep{Hut1981,Leconte2010,Barnes2017}. Assuming zero obliquity ($\psi = 0$) and zero eccentricity ($e = 0$), it yields
\begin{equation}
\frac{d\omega}{dt} = \frac{3 k_2}{\alpha M_\mathrm{pl}} \frac{G^2 M_\star^2 (M_\star + M_\mathrm{pl}) r_\mathrm{pl}^3}{n r_\mathrm{orb}^6} \tau \left( 1 - \frac{\omega}{n} \right), \label{CTL_gov}
\end{equation}
where $\omega$ is the planetary rotation rate, $k_2$ is the second-degree Love number, $\alpha = I/(M_\mathrm{pl} r_\mathrm{pl}^2)$, $I$ is the moment of inertia, $M_\mathrm{pl}$ is the mass of the planet, $\tau$ is the tidal time lag, and $n$ is the mean orbital motion, given by
\begin{equation*}
     n = \frac{v_\mathrm{orb}}{r_\mathrm{orb}} = \sqrt{\frac{G M_\star}{r_\mathrm{orb}^3}}.
\end{equation*}
Replacing $n^2 r_\mathrm{orb}^3 = G(M_\star + M_\mathrm{pl})$ in Eq.~(\ref{CTL_gov}), it yields
\begin{equation}
    \frac{d \omega}{dt} = - \frac{3 k_2}{\alpha} \frac{G M_\star^2 r_\mathrm{pl}^3}{M_\mathrm{pl} r_\mathrm{orb}^6} \tau (\omega - n). \label{CTL_gov_final}
\end{equation}
Assuming constant planetary parameters and a fixed orbital distance, the solution is
\begin{equation}
    \omega(t) = n + (\omega_\mathrm{i} -n ) \exp\left( - \frac{t}{\tau_\mathrm{lock}}  \right). \label{omega_full}
\end{equation}
where $\omega_\mathrm{i}$ is the initial rotation rate, and
\begin{equation}
    \tau_\mathrm{lock} = \frac{1}{\tau} 
    \frac{\alpha}{3 k_2} \frac{M_\mathrm{pl} r_\mathrm{orb}^6}{G M_\star^2 r_\mathrm{pl}^3}
\end{equation}
is the characteristic tidal synchronization timescale.

\subsection{Dynamo scaling}
Earth's magnetic field is generated by a self-sustained dynamo hosted in its outer liquid iron core \citep{1978mfge.book.....M}. This dynamo is strongly influenced by Coriolis forces due to planetary rotation \citep[][]{Christensen2011}. As the final synchronized rotation rate can be significantly lower than the initial rotation rate ($\omega_\mathrm{i}$), the planetary dynamo is expected to be directly affected by tidal locking. A straightforward method to quantify this effect is through the use of dynamo scaling laws, which are widely used in planetary contexts \citep[e.g.][]{2013GeoJI.195...67D}. While a wide variety of scaling laws have been proposed \citep[see][and references therein]{Christensen2010}, here we focus on two main approaches: rotation-based and energy-flux-based scalings.

\renewcommand{\arraystretch}{1.1} 
\begin{table*}[h!]
\centering
\caption{Stellar parameters from the theoretical grid}
\begin{tabular}{cccccccccc}
\hline\hline
SpT & $M_\star~[M_\odot]$ & $R_\star~[R_\odot]$ & $\log (L_\star/L_\odot)$ & Regime & $P_\mathrm{rot}~[\mathrm{days}]$ & $\mathrm{Ro}$ & $B_0~[\mathrm{G}]$ & $t_\star ~[\mathrm{Gyr}]$ & HZ [AU] \\ 
\hline
M6.5 & 0.093 & 0.126 & -3.10 & S & 24.39 & 0.156 & 400 & $2.82^{+0.37}_{-0.33}$ & $(0.02) 0.03, 0.06$ \\
M4.5 & 0.184 & 0.217 & -2.40 & S & 11.67 & 0.101 & 400 & $1.35^{+0.12}_{-0.11}$ & $(0.05) 0.07, 0.13$ \\
M3.5 & 0.27 & 0.3 & -2.03 & S & 7.72 & 0.087 & 400 & $0.35^{+0.02}_{-0.02}$ & $(0.08) 0.10, 0.20$ \\
M1.5 & 0.47 & 0.482 & -1.44 & S & 4.24 & 0.086 & 50 & $0.17^{+0.02}_{-0.01}$ & $(0.15) 0.20, 0.38$ \\
M0.0 & 0.57 & 0.588 & -1.16 & S & 3.44 & 0.091 & 50 & $0.15^{+0.01}_{-0.01}$ & $(0.20) 0.27, 0.52$ \\
\hline
M6.5 & 0.093 & 0.126 & -3.10 & NS & 78.01 & 0.500 & 89 & $4.81^{+2.42}_{-1.61}$ & $(0.02) 0.03, 0.06$ \\
M4.5 & 0.184 & 0.217 & -2.40 & NS & 58.01 & 0.500 & 89 & $4.02^{+1.40}_{-1.04}$ & $(0.05) 0.07, 0.13$ \\
M3.5 & 0.27 & 0.3 & -2.03 & NS & 44.32 & 0.500 & 89 & $3.53^{+0.57}_{-0.49}$ & $(0.08) 0.10, 0.20$ \\
M1.5 & 0.47 & 0.482 & -1.44 & NS & 24.69 & 0.500 & 11 & $2.67^{+0.55}_{-0.46}$ & $(0.15) 0.20, 0.38$ \\
M0.0 & 0.57 & 0.588 & -1.16 & NS & 18.82 & 0.500 & 11 & $1.33^{+0.19}_{-0.17}$ & $(0.20) 0.27, 0.52$ \\
\hline
\end{tabular} 
\tablefoot{From left to right: spectral type, stellar mass $M_\star$, stellar radius $R_\star$, luminosity $L_\star$, regime: saturated (S) or unsaturated (NS), rotation period $P_\mathrm{rot}$, global Rossby number Ro, average surface magnetic field $B_0$, stellar age $t_\star$, and habitable zone $(\ri^\mathrm{TL})\ri,\ro$.}
\label{table-stellar}
\end{table*}

\subsubsection{Model A}
In this model, we adopt scaling laws based on rotation. In particular, those proposed by \cite{1976PEPI...12..350B} and \cite{1993JGG....45...65S}
 \begin{align*}
     B_\mathrm{pl}^2 &\propto \rho_\mathrm{c} \omega^2 R_\mathrm{c}^2, & B_\mathrm{pl}^2 &\propto \rho_\mathrm{c} \omega^2 R_\mathrm{c},
 \end{align*}
where $R_\mathrm{c}$ and $\rho_\mathrm{c}$ are the radius and density of the core. As $B_\mathrm{pl} \propto \omega$, the planetary magnetic field at the pole influenced by rotation is expressed as 
 \begin{equation}
    B_\mathrm{p,0}^{(\omega)}(t) = (\omega(t)/\omega_\mathrm{i}) B_\mathrm{p,0}, \label{modelA}
 \end{equation}
where the rotation rate is given by Eq.~(\ref{omega_full}).

\begin{figure}[t!]
    \centering
    \includegraphics[width=\hsize]{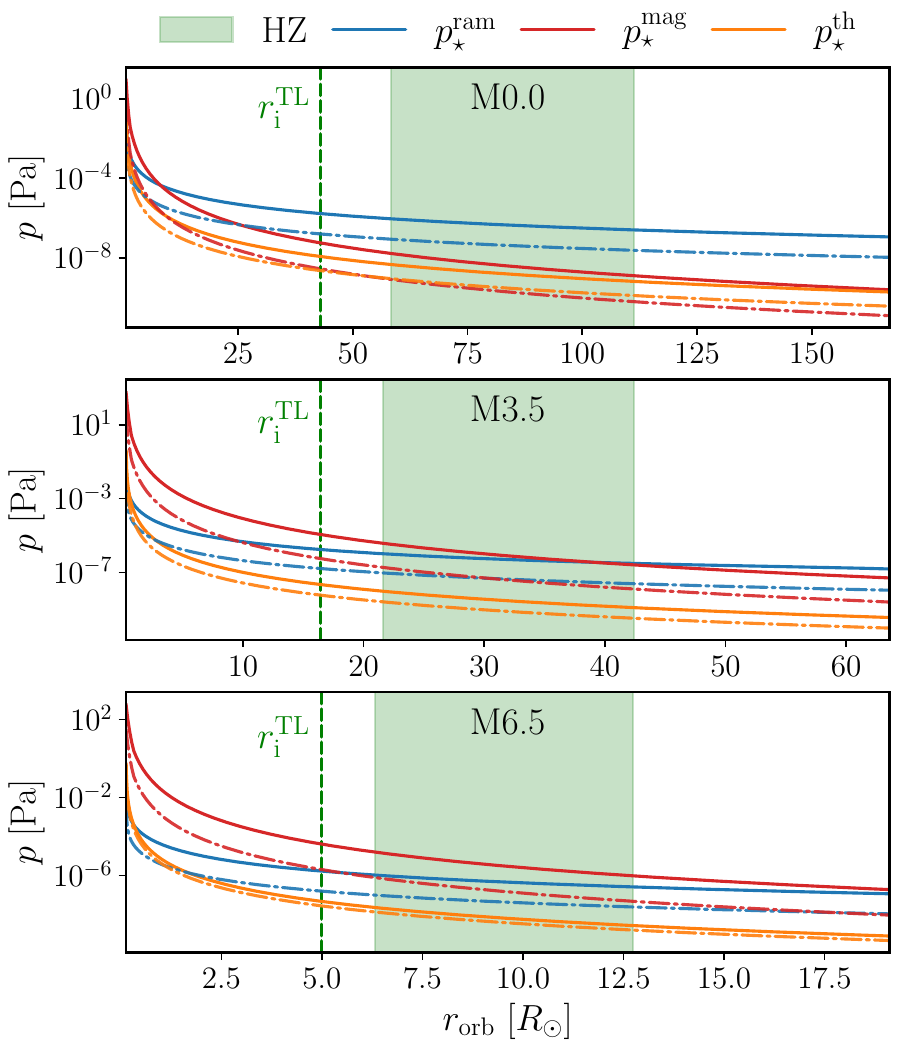}
    \caption{Ram $p_\star^\mathrm{ram}$, magnetic $p_\star^\mathrm{mag}$ and thermal $p_\star^\mathrm{th}$ pressures from the M0.0, M3.5 and M6.5 systems of our theoretical sample. The solid lines represent the saturated cases, and the dash-dotted lines the unsaturated cases. The HZ is given by $(\ri,\ro)$ and $\ri^\mathrm{TL}$ is represented as a vertical dashed line.}
    \label{fig:p_SpT}
\end{figure}

\begin{table}[t!]
\centering
\caption{Initial planetary parameters.}
\label{tab:planet_parameters}
\begin{tabular}{lcc}
\hline \hline
Parameter  & Modern Earth (ME) & Early Earth (EE) \\ \hline
$P_{\mathrm{rot,i}}$ & 24 h & 13 h \\
$\omega_\mathrm{i}$ & $7.272 \times 10^{-5}$ rad/s & $1.343 \times 10^{-4}$ rad/s \\
$B_\mathrm{p,0}$ & 1.0 G & 0.7 G \\
$\mathrm{Ro}_{\ell,0}$ & 0.09 & 0.04 \\
$\tau$ & 640 s & 37 s \\
$Q$ & $\sim 12$ & $\sim 100$ \\
$k_2$ & 0.30 & 0.35 \\\hline
\end{tabular}
\tablefoot{From top to bottom: Initial rotation period $P_\mathrm{rot,i}$, initial rotation rate $\omega_\mathrm{i} = 2\pi/P_\mathrm{rot,i}$, unsuppressed polar field $B_\mathrm{p,0}$, initial local Rossby number $\mathrm{Ro}_\mathrm{\ell,0}$, tidal time lag $\tau$, tidal dissipation factor $Q$ and second-degree Love number $k_2$.}
\end{table}

\renewcommand{\arraystretch}{1.1} 
\begin{table*}[t!]
\centering
\caption{Magnetospheres and magnetic fields of planets at the HZ.}
\begin{tabular}{cc|cc|cc}
\hline\hline
\multicolumn{2}{c}{M dwarfs} & \multicolumn{2}{c}{Model A} & \multicolumn{2}{c}{Model B}
\\
\hline
& &  \multicolumn{4}{c}{Modern-Earth at $(\ri^\mathrm{TL})\ri, \ro$} 
\\
SpT & Regime & $r_\mathrm{M}/r_\mathrm{pl}$ & $B_\mathrm{p,0}^{(\omega)}$ [G] & $r_\mathrm{M}/r_\mathrm{pl}$ & $B_\mathrm{p,0}^{(\omega)}$ [G] \\ 
\hline
M6.5 & S & $(1.38 ) 1.43  , 1.56 $ &  $ (0.24 )  0.17 , 0.06 $  & $(1.04 ) 1.21  , 1.87 $ &  $ (0.10 )  0.10 , 0.10 $ \\
M4.5 & S & $(1.21 ) 1.25  , 1.32 $ &  $ (0.10 )  0.07 , 0.03 $  & $(1.20 ) 1.41  , 2.09 $ &  $ (0.10 )  0.10 , 0.10 $ \\
M3.5 & S & $(1.11 ) 1.14  , 1.17 $ &  $ (0.07 )  0.04 , 0.02 $  & $(1.27 ) 1.49  , 2.14 $ &  $ (0.10 )  0.10 , 0.10 $ \\
M1.5 & S & $(1.22 ) 1.16  , 4.19^{+0.08}_{-0.09}$ &  $ (0.03 )  0.02 , 0.51^{+0.03}_{-0.03}$  & $(1.76 ) 1.94  , 2.43 $ &  $ (0.10 )  0.10 , 0.10 $ \\
M0.0 & S & $(1.09 ) 1.06^{+0.02}_{-0.01} , 5.02^{+0.02}_{-0.02}$ &  $ (0.02 )  0.02 , 0.87^{+0.01}_{-0.01}$  & $(1.77 ) 1.96  , 5.17^{+0.02}_{-0.03}$ &  $ (0.10 )  0.10 , 0.95^{+0.01}_{-0.02} $ \\
M6.5 & NS & $(2.26 ) 2.34  , 2.49 $ &  $ (0.24 )  0.16 , 0.06 $  & $(1.70 ) 1.98  , 2.99 $ &  $ (0.10 )  0.10 , 0.10 $ \\
M4.5 & NS & $(1.97 ) 2.02  , 2.07 $ &  $ (0.10 )  0.07 , 0.03 $  & $(1.95 ) 2.28  , 3.28 $ &  $ (0.10 )  0.10 , 0.10 $ \\
M3.5 & NS & $(1.80 ) 1.83  , 1.83 $ &  $ (0.07 )  0.04 , 0.02 $  & $(2.05 ) 2.40  , 3.36 $ &  $ (0.10 )  0.10 , 0.10 $ \\
M1.5 & NS & $(1.81 ) 1.73  , 1.55^{+0.01}_{-0.00}$ &  $ (0.03 )  0.02 , 0.01 $  & $(2.61 ) 2.88  , 3.60 $ &  $ (0.10 )  0.10 , 0.10 $ \\
M0.0 & NS & $(1.61 ) 1.54  , 5.15^{+0.28}_{-0.30}$ &  $ (0.02)  0.01 , 0.29^{+0.05}_{-0.05}$  & $(2.63 ) 2.91  , 3.62 $ &  $ (0.10 )  0.10 , 0.10 $ \\
\hline
& &  \multicolumn{4}{c}{Early-Earth at $(\ri^\mathrm{TL})\ri, \ro$} 
\\
SpT & Regime & $r_\mathrm{M}/r_\mathrm{pl}$ & $B_\mathrm{p,0}^{(\omega)}$ [G] & $r_\mathrm{M}/r_\mathrm{pl}$ & $B_\mathrm{p,0}^{(\omega)}$ [G] \\ 
\hline
M6.5 & S & $(1.00 ) 1.04  , 1.13 $ &  $(0.09 )  0.06 , 0.02$  & $(0.92 ) 1.07  , 1.66 $ &  $(0.07 )  0.07 , 0.07$ \\
M4.5 & S & $(0.88 ) 0.91  , 0.95 $ &  $(0.04 )  0.03 , 0.01$  & $(1.07 ) 1.26  , 1.86 $ &  $(0.07 )  0.07 , 0.07$ \\
M3.5 & S & $(0.80 ) 0.83  , 2.40^{+0.06}_{-0.06}$ &  $ (0.03 )  0.02 , 0.14^{+0.01}_{-0.01}$  & $(1.12 ) 1.33  , 1.90 $ &  $(0.07 )  0.07 , 0.07$ \\
M1.5 & S & $(0.88 ) 1.76^{+0.11}_{-0.11} , 4.57^{+0.01}_{-0.01}$ &  $(0.01 )  0.07^{+0.01}_{-0.01}, 0.67$  & $(1.56 ) 1.73  , 4.64 $ &  $(0.07 )  0.07 , 0.70$ \\
M0.0 & S & $(1.41^{+0.10}_{-0.10}) 3.23^{+0.04}_{-0.05} , 4.65 $ &  $(0.05^{+0.01}_{-0.01})  0.45^{+0.02}_{-0.02}, 0.69$  & $(1.57 ) 3.76  , 4.67 $ &  $(0.07 )  0.70 , 0.70$ \\
M6.5 & NS & $(1.64 ) 1.69  , 1.81 $ &  $(0.09 )  0.06 , 0.02$  & $(1.51 ) 1.76  , 2.66 $ &  $(0.07 )  0.07 , 0.07$ \\
M4.5 & NS & $(1.43 ) 1.46  , 1.50 $ &  $(0.04 )  0.03 , 0.01$  & $(1.73 ) 2.03  , 2.91 $ &  $(0.07 )  0.07 , 0.07$ \\
M3.5 & NS & $(1.30 ) 1.33  , 1.33 $ &  $(0.03 )  0.02 , 0.01$  & $(1.82 ) 2.13  , 2.98 $ &  $(0.07 )  0.07 , 0.07$ \\
M1.5 & NS & $(1.31 ) 1.25  , 5.41^{+0.23}_{-0.26}$ &  $ (0.01 )  0.01 , 0.34^{+0.04}_{-0.05}$  & $(2.32 ) 2.56  , 6.83^{+0.06}_{-1.48}$ &  $ (0.07 )  0.07 , 0.68^{+0.02}_{-0.35}$ \\
M0.0 & NS & $(1.17 ) 1.62^{+0.22}_{-0.18} , 6.73^{+0.02}_{-0.03}$ &  $ (0.01 )  0.02^{+0.01}_{-0.01}, 0.64^{+0.01}_{-0.01}$  & $(2.33 ) 2.59  , 6.92 $ &  $(0.07 )  0.07 , 0.70$ \\
\hline
\end{tabular} 
\tablefoot{The planetary magnetic field at the pole influenced by rotation $B_\mathrm{p,0}^{(\omega)}$ and the upstream magnetospheres $r_\mathrm{M}/r_\mathrm{pl}$ are estimated for Modern and Early Earth-like planets using Model A and Model B. The displayed values correspond to the planets orbiting at $(\ri^\mathrm{TL})\ri, \ro$. In the tidally locked, and the (almost) freely-rotating cases, the errors are negligible and are not included.}
\label{table-planet}
\end{table*}

\subsubsection{Model B \label{modelB}}
Direct rotation scalings such as Model A were used by \cite{Griessmeier2004, Griessmeier2005, Griessmeier2009}, which led to an enormous weakening of the planetary magnetic field. Interestingly, in energy-flux-based scalings, the magnetic field strength depends mainly on the convective
energy flux available in the core of the planet, and it is independent of the rotation rate \citep{Christensen2011}. Scaling laws by \cite{Christensen2006} suggest that while planetary magnetic field strength is primarily driven by the convective heat flux ($B_\mathrm{pl} \propto q_\mathrm{c}^{1/3}$), rotation plays a critical role in shaping the geometry of it. The transition between a dipolar field and a multipolar topology is thought to be governed by the balance between inertial and Coriolis forces, often parameterized by the local Rossby number $\mathrm{Ro}_\ell = \overline{\ell}/\pi\,\mathrm{Ro} $, where $\overline{\ell}$ is the characteristic length scale of the flows \citep{Christensen2006, Olson2006, Gastine2013}. In regimes where Coriolis forces dominate ($\mathrm{Ro}_\ell \lesssim 0.1$), stable dipolar solutions emerge, whereas higher inertial contributions lead to multipolar configurations \citep{Gastine2012}. Although other factors such as density stratification \citep{Zaire2022}, viscous forces \citep{Soderlund2012} and the Lorentz force \citep{Menu2020, Hidalgo2025a} may influence this threshold, the local Rossby number remains a key diagnostic for the magnetic topologies of rotating dynamos. 

Following \cite{Olson2006}, the local Rossby number scales with $\mathrm{Ro}_\ell \propto \omega^{-7/6}$, and other quantities that essentially do not change over the course of the rotational spin-down. Therefore
\begin{equation}
    \mathrm{Ro}_\ell (t) \approx \mathrm{Ro}_\mathrm{\ell,0} \left( \frac{\omega(t)}{\omega_\mathrm{i}} \right)^{-7/6}, \label{Ro_l}
\end{equation}
where $\mathrm{Ro}_\mathrm{\ell,0}$ is the initial local Rossby number of the planet.

Numerical simulations show that the boundary between the stable dipolar and multipolar regimes is typically inherently sharp \citep{Christensen2006}. To capture this rapid transition, we use a “soft step function” approach, similar to that of \cite{Zuluaga2013}, where the resulting magnetic field at the pole is given by $B_\mathrm{p,0}^{(\omega)} = f_\mathrm{scale} B_\mathrm{p,0}$, with
\begin{equation}
    f_\mathrm{scale} = f_\mathrm{dip,TL} + \frac{f_\mathrm{dip} - f_\mathrm{dip,TL}}{1 + e^{k(\mathrm{Ro}_\ell - \mathrm{Ro}_\mathrm{\ell,crit})}},
\end{equation}
where $f_\mathrm{dip}$ is the initial dipolarity, assumed to be 1 (see Eq.~\ref{planetB}), $f_\mathrm{dip,TL}$ is the residual dipolar fraction after the topology transition, $\mathrm{Ro}_\mathrm{\ell,crit}$ is the critical Rossby number of the transition, fixed at 0.12 \citep{Olson2006}, and $k$ is a steepness parameter controlling the transition width, fixed at $k=200$ to prevent the artificial reduction of planetary magnetic fields deep within the stable dipolar regime. Moreover, although multipolar dynamos are typically classified as $f_\mathrm{dip} < 0.35$ \citep{Christensen2010, Zuluaga2012}, their final dipolarity can  span several orders of magnitude, dropping as low as $f_\mathrm{dip}\approx 10^{-4}$ in certain configurations \citep{Gastine2012}. Given this intrinsic dispersion, we adopt $f_\mathrm{dip,TL} = 0.1 f_\mathrm{dip}$ as a representative, conservative baseline, that lies comfortably within the established theoretical bounds \citep[e.g.][]{Olson2006}. Additionally, because the tidal locking timescale is significantly shorter than the planetary secular cooling timescale, and terrestrial cores maintain a remarkably steady convective energy flux over gigayear timescales \citep[e.g.,][]{Zuluaga2013}, we assume a fixed $q_\mathrm{c}$ for all modeled planets.

\section{Results \label{results}}

In Fig.~\ref{fig:illu}, we provide a summary of the physical processes we aim to explore here, particularly the impact of tidal locking on the rotation period of the planets, the local Rossby number and the dynamo topology, as well as the direct consequences for planetary magnetic protection. To explore the predictions of our framework and to have a better control of the parameter space, we apply our model to a theoretical grid of M dwarfs with five different spectral types. Additionally, in Section~\ref{real-examples} we use our model in a sample of observed stars with masses between $0.10-0.75~M_\odot$.

\begin{figure*}[t!]
    \centering
    \includegraphics[scale=0.5]{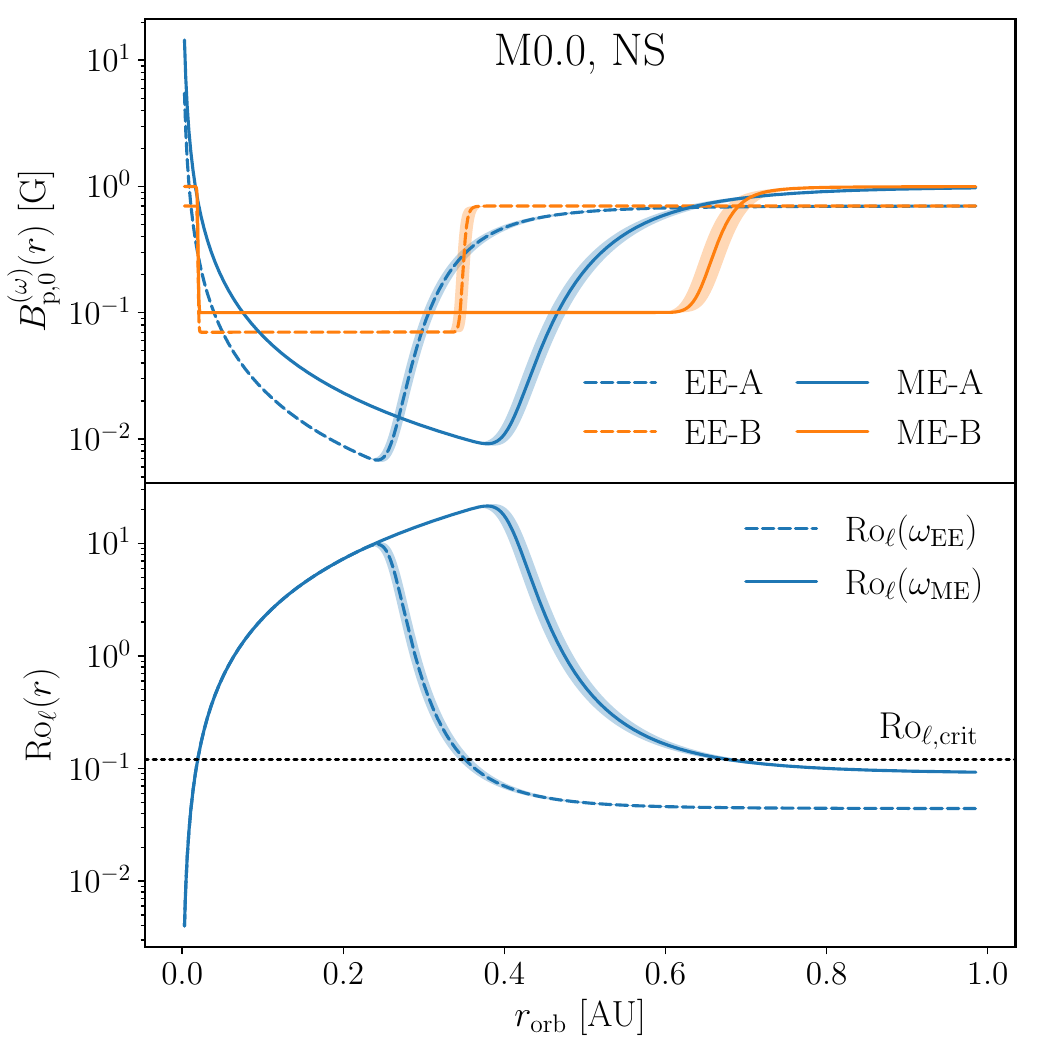}
    \includegraphics[scale=0.5]{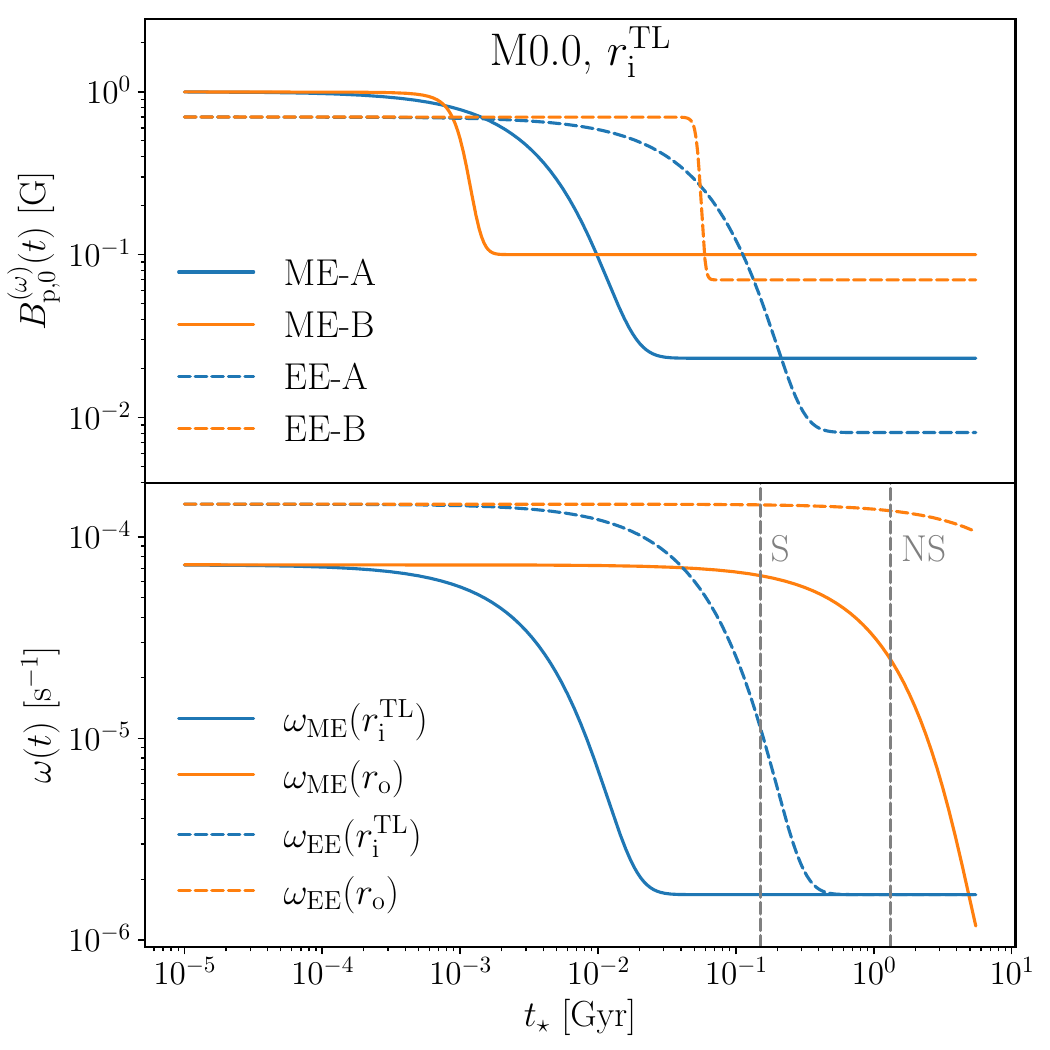}
    \caption{Spatial dependence and temporal evolution of magnetic and rotational parameters for planets orbiting an M0.0 host star. \textit{Left panels:} Planetary magnetic field (\textit{top}) and local Rossby number (\textit{bottom}) as functions of orbital distance in the unsaturated (NS) regime. The filled areas are the propagated uncertainties. \textit{Right panels:} Planetary magnetic field at $r_\mathrm{i}^\mathrm{TL}$ (\textit{top}) and rotation rate at $r_\mathrm{i}^\mathrm{TL}$ and $r_\mathrm{o}$ (\textit{bottom}) as functions of time. The vertical lines indicate the stellar ages corresponding to the saturated (S) and unsaturated (NS) regimes.
    }

    \label{fig:B0P}
\end{figure*}

\subsection{Stellar wind environments \label{wind}}

For our theoretical sample of M dwarfs, we adopt stellar masses from $0.093~M_\odot$ to $0.57~M_\odot$, which corresponds to spectral types from M6.5 to M0.0. To cover the vast observational diversity of M-dwarf rotation periods and magnetic fields \citep[e.g.][]{Kochukhov2021}, we consider two distinct evolutionary states for each mass: a saturated case and an unsaturated case. In both regimes, the convective turnover time $\tau_\mathrm{conv}$ is calculated using the mass-dependent relations of \cite{Wright2018}. In the saturated regime, the rotation period $P_\mathrm{rot}^\mathrm{sat} = 2\pi/ \Omega_\mathrm{sat}$ is determined with the rotation rate given by Eq.~(\ref{omega-sat}), while its corresponding Rossby number is derived as $\mathrm{Ro}_\mathrm{sat} = P_\mathrm{rot}^\mathrm{sat}/\tau_\mathrm{conv}$. For the unsaturated regime, it yields $P_\mathrm{rot}^\mathrm{unsat} = \mathrm{Ro}_\mathrm{unsat} \tau_\mathrm{conv}$, where we adopt a quiet-state baseline of $\mathrm{Ro}_\mathrm{unsat} = 0.5$, following \cite{RodriguezMozos2019}. We assume that the base surface magnetic field $B_0$ is roughly equivalent to the ZDI large-scale magnetic field, i.e. $B_0 \approx \langle |B_V| \rangle$. Following the empirical trends reported by \cite{Vidotto2014b}, these observed large-scale magnetic fields tend to converge toward two distinct limits in the saturated regime based on the spectral type, we set $B_0 = 10^{2.6}~\mathrm{G} \approx 400~\mathrm{G}$ for mid-to-late M dwarfs (M3.5, M4.5, and M6.5) and $10^{1.7}~\mathrm{G} \approx 50~\mathrm{G}$ for early M dwarfs (M0.0 and M1.5). For unsaturated cases, $B_0$ is scaled as $P_\mathrm{rot}^{-1.32}$ and fixed to these saturation values at the critical Rossby threshold $\mathrm{Ro}_\mathrm{sat}^\mathrm{crit} = 0.16$ \citep{Wright2011}. For the stellar radii and luminosities, we use a standard value based on the spectral class following \cite{Pecaut2013}\footnote{The standard values for main-sequence stars are provided in this \href{https://www.pas.rochester.edu/~emamajek/EEM_dwarf_UBVIJHK_colors_Teff.txt}{table}.}. Finally, to estimate the stellar age we use the rotation-dependent empirical relations of \cite{Engle2023}. It should be noted that the stellar mass and $\mathrm{Ro}_\mathrm{unsat}$ are the only independent input parameters in our framework, and our saturated systems represent  the oldest age and longest rotation period at which the host star maintains the magnetic saturation. The resulting stellar parameters of our sample are displayed in Table~\ref{table-stellar}.

In Figure~\ref{fig:p_SpT}, the radial profile of the stellar pressures from representative cases are shown. The unsaturated cases exhibit lower pressures, mainly due to reduced rotation and magnetic flux. The ram pressures between both cases differ by roughly a factor of 10 in the HZ of all the models. A similar situation happens with the magnetic pressure, where the ratio between saturated and unsaturated is $\sim 20$. In the thermal pressure the difference is minor, where the ratios are $5$ (M0.0), $3.75$ (M3.5) and $1.7$ (M6.5). In the M0.0 models, the ram pressure dominates over the rest of contributions in the HZ, and the region where the magnetic pressure is the strongest is significantly close to the star. In M3.5, the magnetic pressure dominates in the inner radius of the HZ. However, at some point inside the HZ the ram pressure overcomes the magnetic pressure. This transition happens closer to $\ro$ in the saturated case, and around the center in the unsaturated case. Finally, in the M6.5 models, the magnetic pressure is stronger in the entire extension of the HZ.

\begin{figure*}[h!]
    \centering
    \includegraphics[width=\hsize]{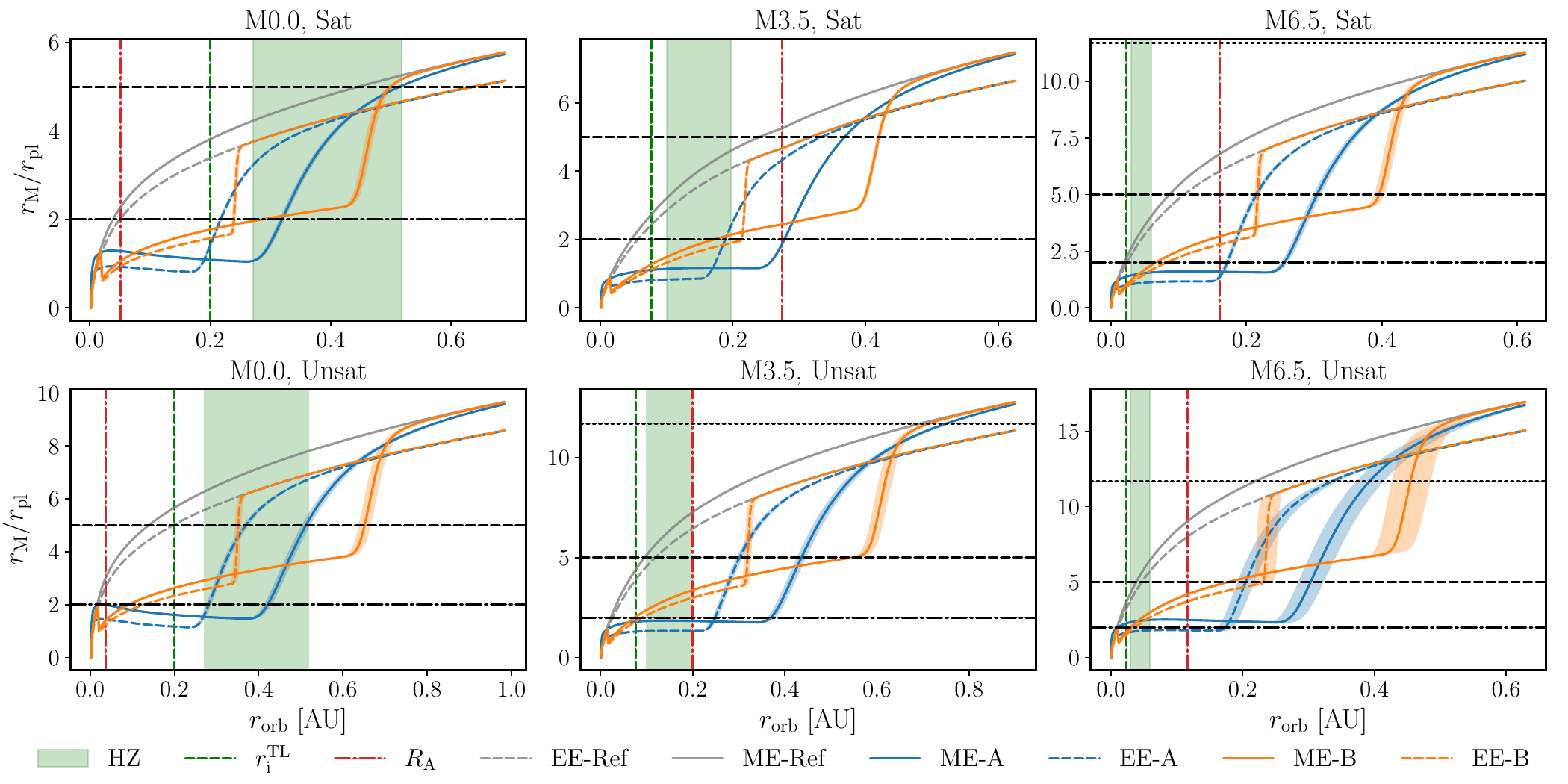}
    \caption{The size of the upstream magnetosphere as a function of the orbital distance from representative cases of our theoretical systems. Solid lines denote Modern Earth (ME) planets, while dashed lines indicate Early Earth (EE) planets. The filled areas are the propagated uncertainties. The gray lines (Ref) represent the case where tidal locking is not considered. The vertical red line is the Alfvén Radius $R_\mathrm{A}$. The horizontal lines represent magnetospheres $r_\mathrm{M}/r_\mathrm{pl}$ of 2 (dashdot), 5 (dashed) and 11.7 (dotted).}
    \label{fig:rmp_SpT}
\end{figure*}

\subsection{Magnetospheric protection of Earth-like planets}

To model the hypothetical planets, we establish two distinct baselines modeled after different epochs of Earth: a Modern Earth (ME) scenario and a Paleoarchean Early Earth (EE) analogue. For both scenarios, we adopt standard terrestrial structural parameters, keeping the planetary radius fixed at $r_\mathrm{pl} = 6378$ km, the mass at $M_\mathrm{pl} = 5.972 \times 10^{24}$ kg and $\alpha = 0.3$ \citep{Griessmeier2009}. The specific rotational, magnetic, and tidal parameters configured for each baseline model are summarized in Table~\ref{tab:planet_parameters}.

In ME we fixed the local Rossby number at $\mathrm{Ro}_{\ell,\mathrm{ref}} = 0.09$ \citep{Olson2006} and the polar field strength at $B_\mathrm{p,0} = 1.0$ G \citep{Vidotto2013}. To model its tidal evolution within the CTL framework, we adopt a constant time lag of $\tau = 640$ s \citep{Heller2011, Barnes2017}, which is consistent with the unusually low modern terrestrial tidal dissipation factor of $Q = 12 \pm 2$ \citep{Williams1978, Hubbard1984} and $k_2 = 0.3$ \citep{Yoder1995, Barnes2017}. During the Paleoarchean, the Solar day is estimated to be $\sim 13$h \citep{Eulenfeld2023}. Due to this rapid rotation, the initial local Rossby number scales down to a baseline of $\mathrm{Ro}_{\ell,0} \approx 0.04$. The baseline polar field strength is set to a conservative $B_\mathrm{p,0} = 0.7$ G, in accordance with Paleoarchean constraints \citep{Tarduno2010, Tarduno2014}. Given that Earth's $Q$ was likely larger in the geological past, as different continental layout would naturally yield to weaker oceanic dissipation \citep{Green2017}, we set an initial value of $Q \approx 100$, which is also consistent with other terrestrial bodies, like Io and Mars \citep{Seagatz1988,Pou2022}. Although there is no direct conversion between $Q$ and $\tau$, for a non-synchronously rotating object when $e =0$ and $\psi=0$ it yields $\tau \approx (2|\omega - n|Q)^{-1} \approx (2\omega Q)^{-1}$ \citep{Heller2011}, which gives $\tau \approx 37$ s. Finally, we adopt a slightly higher Love number of $k_2 = 0.35$ to account for a hotter, more compliant primitive interior prior to inner core crystallization \citep[see e.g.][]{Farhat2025}.

The upstream magnetospheres $r_\mathrm{M}/r_\mathrm{pl}$ and the magnetic fields $B_\mathrm{p,0}^{(\omega)}$ of planets orbiting the HZ of our stellar sample are shown in Table~\ref{table-planet}. Tidal locking significantly reduces the planetary magnetic fields in most of the cases. In Model A, the reduction is particularly extreme as in some cases $B_\mathrm{p,0}^{(\omega)}$ drops to approximately $1\%$ of its initial value. Notably, at $\ri^\mathrm{TL}$ in ME, Model A (ME-A), we find that the planetary magnetic field is suppressed by a factor of roughly $0.02$ in both saturated and unsaturated M0.0 systems. This finding is in agreement with the work of \cite{Griessmeier2005}, who estimated that an Earth-like planet orbiting a $0.5M_\odot$ star at $r=0.2$ [AU], which fits our value of $\ri^\mathrm{TL}$, reduces its magnetic dipolar moment by a factor between 0.022 and 0.15. While Model A fits with the worst scenario, Model B is closer to their upper limit. In models B, the minimum possible planetary magnetic field is achieved in essentially all the cases at $\ri^\mathrm{TL}$ and $\ri$, except in EE Model B (EE-B) of saturated M0.0. Interestingly, at $r_\mathrm{o}$ of the M1.5 saturated system, the planetary magnetic field under model ME-B has already collapsed to $0.1$ G, whereas in ME-A it has decreased by only $\sim 50\%$. This is a direct consequence of the sharp profile of $f_\mathrm{scale}$, as the dynamo crosses the critical local Rossby number threshold ($\mathrm{Ro}_\ell > \mathrm{Ro}_{\ell, \mathrm{crit}}$) during the intermediate stages of the spin-down. The inherent fragility of a stable dipolar configuration triggers a rapid topological collapse before the planet achieves full synchronization. As visible in Fig.~\ref{fig:B0P}, this behavior is shown in both spatial and temporal dimensions. As a function of distance (left panels), models B fully decay at wider orbital distances than in Model A. Correspondingly, in the temporal evolution (right panels), the topological collapse occurs significantly before the planet reaches full synchronization. It should be noted that very close to the star, $\mathrm{Ro}_\ell$ < $\mathrm{Ro}_\mathrm{\ell, crit}$ (see bottom left panel of Fig.~\ref{fig:B0P}) and the magnetic field becomes dipolar again.  While in Model B the planetary field is strictly constrained between $f_\mathrm{dip}B_\mathrm{p,0}$ and $f_\mathrm{dip,TL}B_\mathrm{p,0}$, in Model A it spans multiple orders of magnitude, going as low as $0.006$~G ($0.004$~G) and exceeding $\sim 60$~G ($22$~G) for ME (EE) orbiting M6.5 systems. However, while these strong fields theoretically provide enhanced magnetic shielding, planets orbiting at such close distances are exposed to harsh stellar wind pressures that effectively compress and neutralize the magnetospheres. As a consequence of the weak planetary magnetic fields, the magnetospheres are also significantly reduced in the tidally locked cases. This is visible in Figure~\ref{fig:rmp_SpT}, where the magnetospheric sizes as functions of the orbital distances are shown.

\subsection{Implications for habitability}

It should be noted that our framework does not directly model atmospheric dynamics or escape mechanisms, therefore, establishing a definitive magnetospheric threshold for habitability is not directly possible within this model. This uncertainty increases in our M3.5, M4.5 and M6.5 systems, where the entire HZ resides in the sub-Alfvénic regime (see Fig.~\ref{fig:rmp_SpT}). In these environments, magnetic star-planet interactions can drive significant atmospheric erosion \citep[e.g.,][]{Laine2012, Strugarek2018}. Nevertheless, we can analyze our results by adopting terrestrial baselines. If a modern Earth-like shield \citep[$r_\mathrm{M}/r_\mathrm{pl} = 11.7$,][]{Vidotto2013} is strictly required for habitability, every planet in our sample would be inhabitable. However, terrestrial atmospheres seem to survive under much higher compression, as was likely the case for Paleoarchean Earth \citep{Tarduno2010}. Rather than imposing a strict, threshold for habitability, we adopt a probabilistic approach to quantify the shielding efficiency of the planetary magnetosphere, following the methodology of \cite{RodriguezMozos2019}. Based on the topology of a dipole field, the fractional area of the planetary atmosphere exposed to the stellar wind is defined by the unprotected polar cap angle, $\alpha_0 = \arcsin\left((r_\mathrm{M}/r_\mathrm{pl})^{-1/2}\right)$ \citep{Vidotto2013}. We define an atmospheric protection likelihood $P(\alpha_0)$ based on a Gaussian decay of this unprotected angle
\begin{equation*}
    P(\alpha_0) = \left\{ \begin{array}{lcc} 1 & \text{if} & \alpha_0 \leq \alpha_\mathrm{0,crit} \\ \exp\left(-\frac{1}{2}  \left[ \frac{\alpha_0 - \alpha_\mathrm{0,crit}}{\sigma} \right]^2 \right)  & \text{if} & \alpha_0 > \alpha_\mathrm{0,crit} \end{array} \right. .
\end{equation*}
where $\alpha_\mathrm{0,crit} \approx 26.6^\circ$ corresponds to a standoff distance of $r_\mathrm{M}/r_\mathrm{pl} = 5$, a regime consistent with Paleoarchean Earth \citep{Tarduno2010}. The decay parameter $\sigma$ is calibrated such that the probability is 0 when $\alpha_0 = 45^\circ$, corresponding to the limit of $r_\mathrm{M}/r_\mathrm{pl} = 2$ proposed by \cite{2007AsBio...7..185L}.

\begin{figure*}[h!]
    \centering
    \includegraphics[width=\hsize]{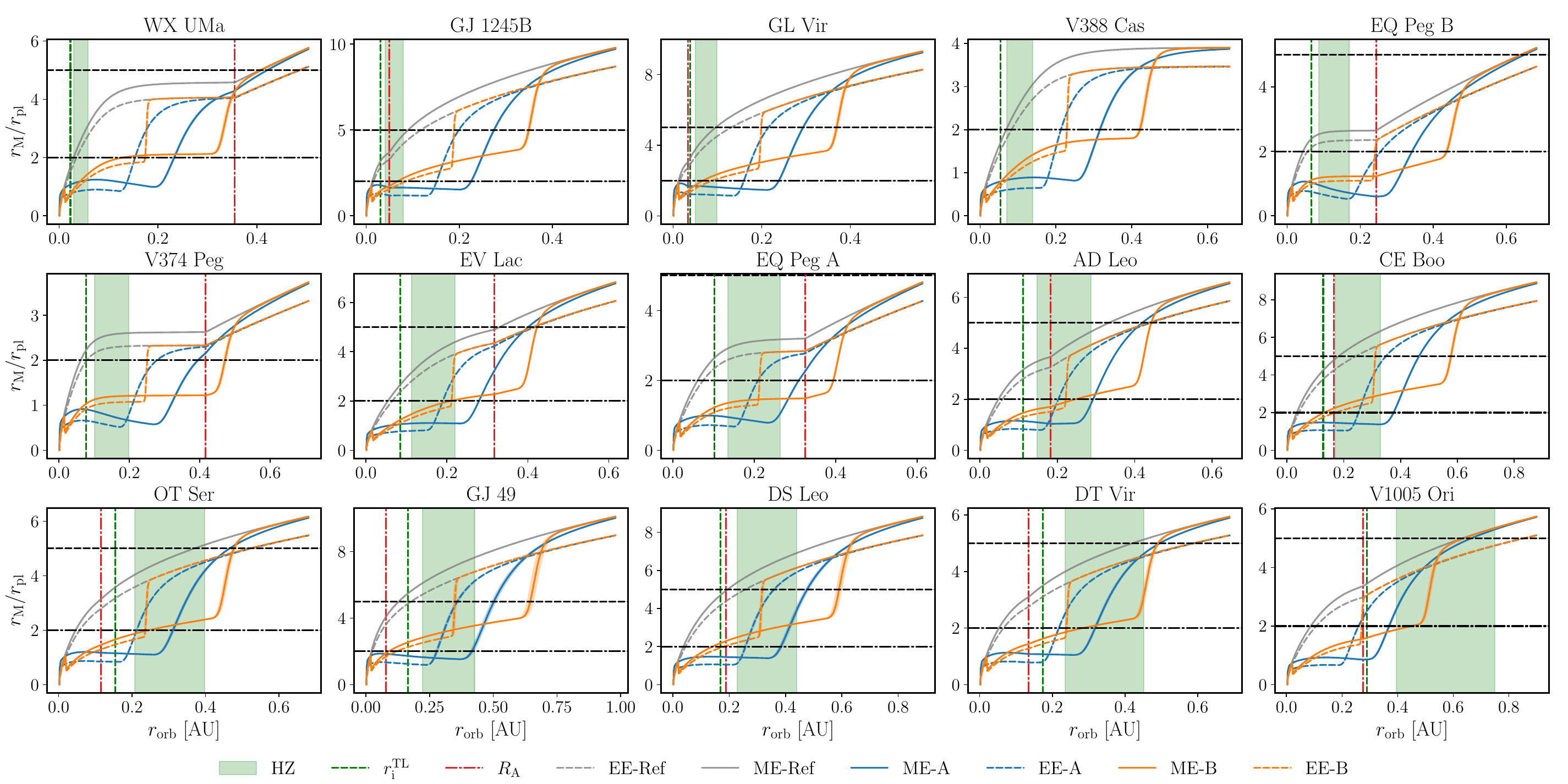}
    \caption{Same as Figure~\ref{fig:rmp_SpT}, but for the sample of 15 M dwarfs described in Section~\ref{real-examples}.}
    \label{fig:rmp_real_cases}
\end{figure*}

\begin{figure*}[h!]
    \centering
    \includegraphics[width=\hsize]{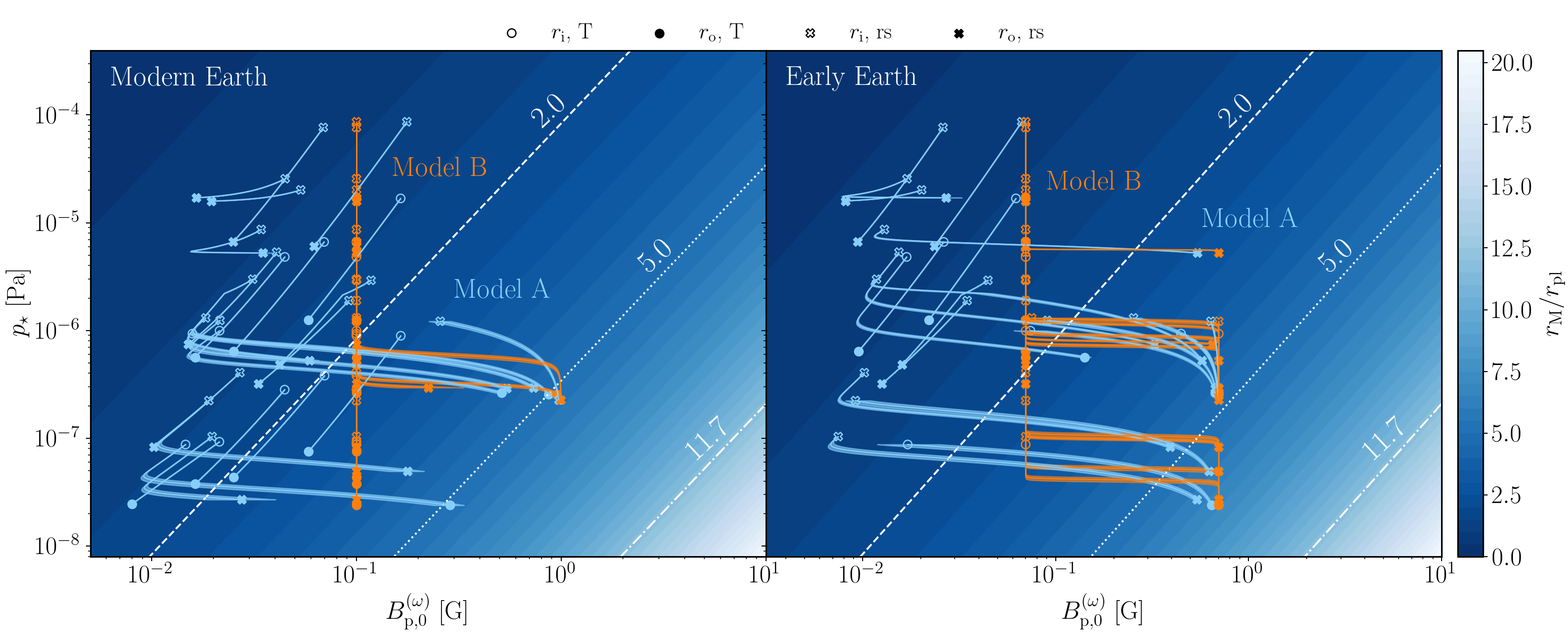}
    \caption{Normalized magnetospheric standoff distance ($r_\mathrm{M}/r_\mathrm{pl}$) as a function of the planetary magnetic field $B_\mathrm{p,0}^{(\omega)}$ and the total stellar pressure $p_\star$ from Modern Earth (\textit{left panel}) and Early Earth (\textit{right panel}) across the HZ of the host star, where the specific markers for the boundaries ($\ri$ and $\ro$) are indicated in the legend. “T” represents the stars from theoretical grid, and “rs” the observed sample of 15 stars.}
    \label{fig:lastplot}
\end{figure*}

As visible in Tables~\ref{table-planet} and \ref{table-probabilities}, magnetic protection at the tidally locked inner boundary ($\ri^\mathrm{TL}$) is severely affected across the entire grid due to rapid rotational synchronization. At this distance, the largest magnetosphere is $r_\mathrm{M}/r_\mathrm{pl} = 2.63$ ($\alpha_0 = 38^\circ$) found in the unsaturated M0.0 system under model ME-B, which yields an atmospheric protection likelihood of only $17\%$. At the traditional inner HZ boundary ($\ri$), the saturated M0.0 system with EE-B offers the highest protection, maintaining a standoff distance of $3.76~r_\mathrm{pl}$ ($\alpha_0 = 31^\circ$) and a $76\%$ likelihood of atmospheric retention. The outer boundary of the HZ ($\ro$) presents the most favorable conditions, particularly for the M0.0 systems, where the planets escape tidal locking, achieving a $99-100\%$ protection likelihood across nearly all scenarios, except for ME-B in the unsaturated system, where the probability drops to $70\%$. After the M0.0 systems, the next highest protection at $\ro$ occurs in the M1.5 systems under EE models, with likelihoods of $98-100\%$. For the lowest-mass systems (M3.5 to M6.5), outer-HZ protection is almost completely neutralized in the saturated systems, and in their unsaturated counterparts only Model B preserves moderate shield efficiencies (e.g., $57\%$ in the unsaturated M3.5 system). It should be noted that the M0.0 and M1.5 systems are the only configurations where the HZ is outside the sub-Alfvénic regime, contributing to the favorable conditions that increase the atmospheric protection likelihood.

\begin{figure*}[h!]
    \centering
    \includegraphics[width=\hsize]{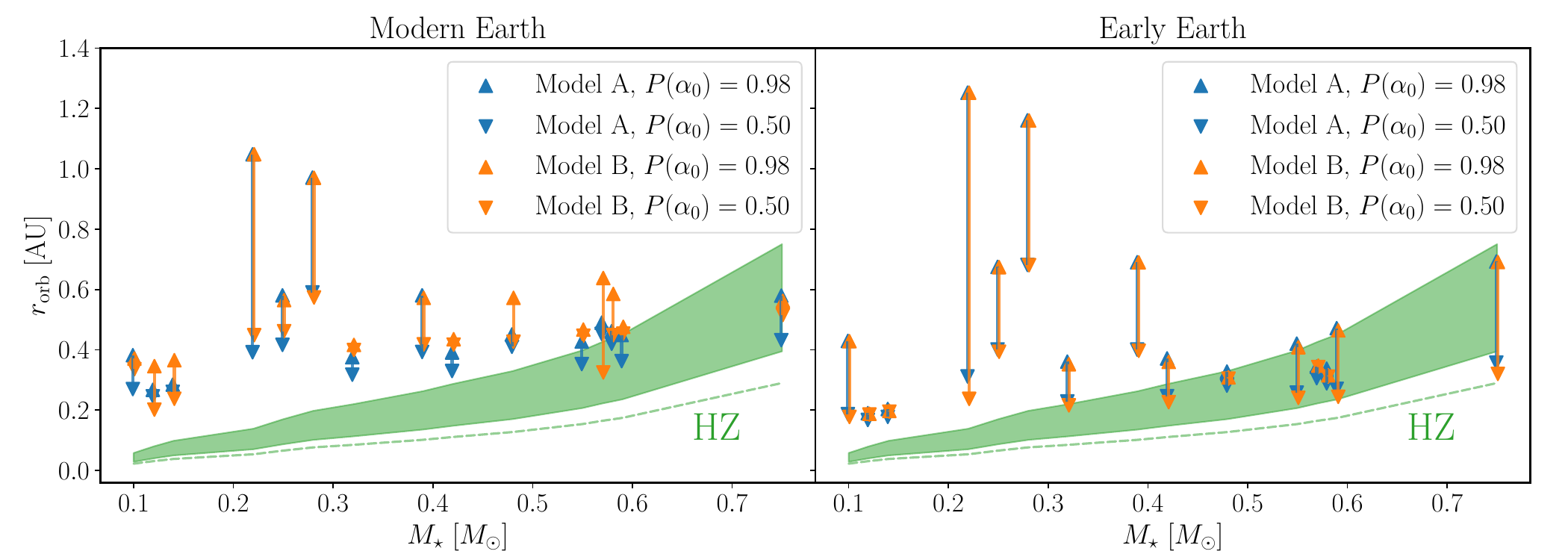}
    \caption{Orbital distance required to achieve an atmospheric protection likelihood of 0.50 (down arrow) and 0.98 (up arrow) as a function of mass, for Modern Earth (\textit{left panel}) and Early Earth (\textit{right panel}) orbiting the sample observed M dwarfs. The green area is the traditional HZ, and the dashed line indicates $\ri^\mathrm{TL}$.}
    \label{fig:like_orb}
\end{figure*}

\subsection{Application to observed systems \label{real-examples}}

In this section, we apply our model to hypothetical planets orbiting the HZs of 15 M dwarfs selected from the literature, with masses ranging between $0.10~M_\odot$ and $0.75~M_\odot$. The stellar parameters of the stars are displayed in Table~\ref{table-stellar-real-cases}. The spectral types, the ZDI magnetic fields (where $B_0 \approx \langle B_V \rangle$ is assumed) and the rotation periods were extracted from Tables 2 and 3 of \cite{Kochukhov2021}. The mass, radius and effective temperature used to infer $L_\star$, originate from the references listed in the last column of Table~\ref{table-stellar-real-cases}. The Rossby number follows the same prescription as in Section~\ref{wind}, and the stellar ages with the corresponding errors are estimated with the empirical relations of \cite{Engle2023}.

Figure~\ref{fig:rmp_real_cases} shows the radial evolution of the upstream magnetospheres for planets orbiting these M dwarfs. In the fast rotators with strong magnetic fields, such as WX UMa, V388 Cas, V374 Peg, EQ Peg A and EQ Peg B, the compressive effects of the sub-Alfvénic regime are substantially more pronounced than in our theoretical grid. It should be noted that in such extreme environments our frameworks can yield magnetospheric standoff distances smaller than the planet itself ($r_\mathrm{M}/r_\mathrm{pl} < 1$). We interpret these cases as the total collapse of the global magnetosphere, where its size becomes vanishingly small. As the magnetic shield is neutralized, allowing the stellar wind to interact directly with the atmosphere, these cases are assigned an atmospheric protection likelihood of zero.

\begin{figure}[h!]
    \centering
    \includegraphics[width=\hsize]{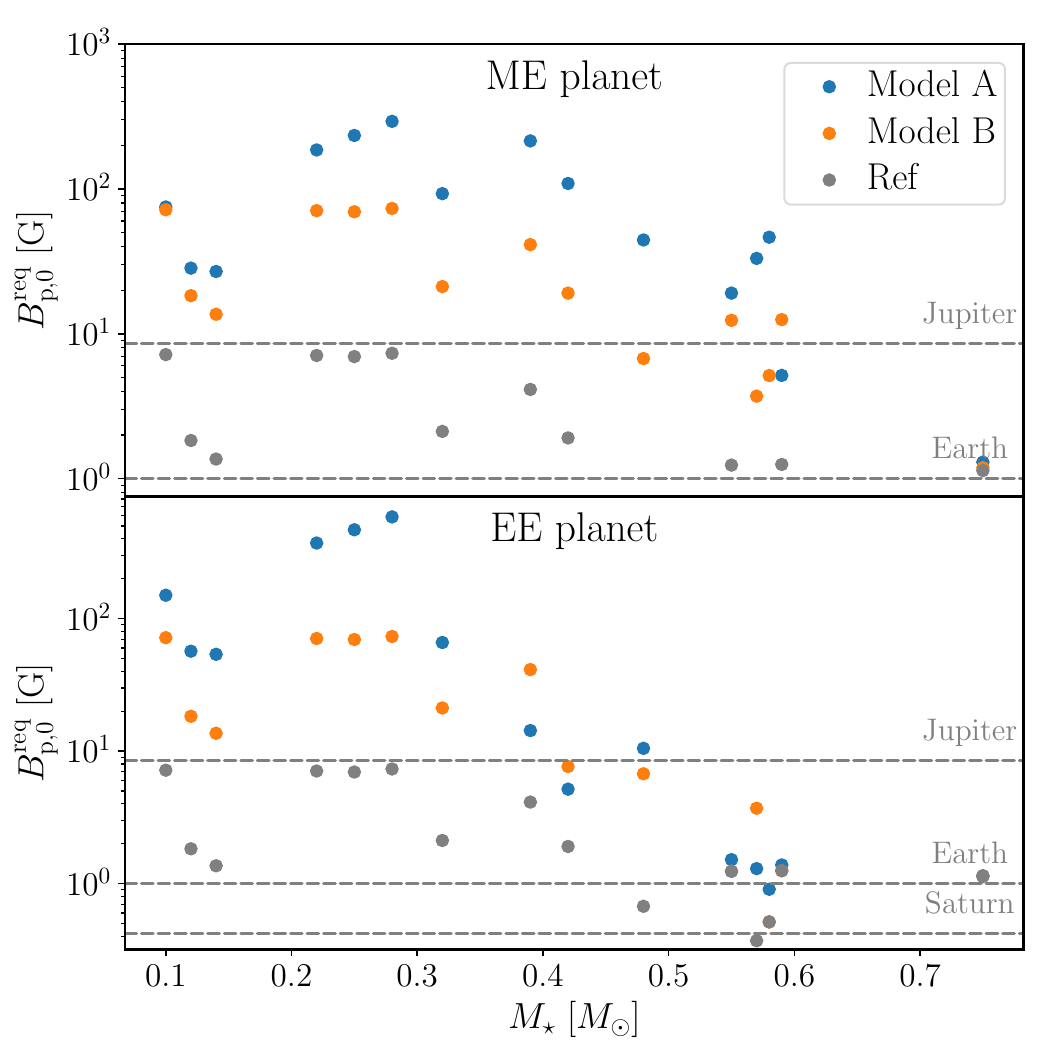}
    \caption{Required initial planetary magnetic field to achieve an atmospheric protection likelihood of $100\%$ at the middle of the HZ, for Modern Earth (\textit{top panel}) and Early Earth (\textit{bottom panel}) planets. The reference values of Jupiter and Saturn are extracted from Table 2 of \cite{RodriguezMozos2019}.}
    \label{fig:req_B0p}
\end{figure}

At the inner habitable zone boundaries ($\ri^\mathrm{TL}$ and $\ri$) most of the planets are tidally locked and their dipolar configuration collapsed, with the notable exception of V1005 Ori for EE, where the dipole remains robust across the entire extent of the HZ. Furthermore, the atmospheric protection likelihood is essentially zero across nearly the entire sample at these close-in orbits. This is visible in Figure~\ref{fig:lastplot}, where most of the sample at $\ri$ is in the $r_\mathrm{M}/r_\mathrm{pl} < 2$ regime. The rare exceptions are found around the few unsaturated stars in our sample. For instance, in the GJ 49 system at $(r_\mathrm{i}^\mathrm{TL}) r_\mathrm{i}$, ME-B predicts a protection likelihood of ($14\%$) $27\%$. This model also predicts a $12\%$ likelihood at $\ri$ for DS Leo, another unsaturated star, which drops to $4\%$ in model EE-B. The V1005 Ori system, presents a favorable inner environment for EE models keeping likelihoods between $64-69\%$ at $\ri$. The only saturated system that offers a reasonable protection at $\ri$ is DT Vir, where the likelihood is $\sim 13\%$ in model EE-A. This prediction is exclusive to this specific configuration, as the remaining models predict a likelihood of essentially zero at such distance.

At the outer boundary of the habitable zone ($\ro$), the EE model exhibits a mass-dependent threshold at $M_\star \approx 0.32~M_\odot$. For lower-mass hosts, rapid tidal synchronization forces surface fields to the $0.07$~G multipolar base, reducing protection likelihoods to typically $\sim 0\%$, with only $1\text{--}2\%$ residual values for GJ 1245B and GL Vir under Model B. At EV Lac, protection likelihoods increase to $38\%$ (EE-A) and $79\%$ (EE-B), reflecting stronger planetary magnetic fields (see Table~\ref{table-real-cases-planet}). For more massive stars ($M_\star \geq 0.48~M_\odot$), outer-HZ protection remains robust, consistently achieving $97-100\%$ likelihoods, even for saturated hosts like OT Ser. On the other hand, the ME scenario lacks a clear mass-dependent transition. Across almost the entire observed sample, the planetary dipole undergoes topological collapse to its $0.10$~G baseline, reducing the likelihoods significantly. Interestingly, for massive stars, the protection likelihoods under Model B remain non-zero even after topological collapse. This is a consequence of their HZs lying outside the sub-Alfvénic regime. Therefore, even a severely weakened magnetic field can offer moderate atmospheric protection against the stellar environment.

These trends are visible in Figure~\ref{fig:like_orb}, where the required orbital distance to achieve protection likelihoods between 0.50 and 0.98 for the Early Earth scenario typically overlaps with the HZ in early M dwarfs, demonstrating that a shielded environment at the outer HZ of M dwarfs requires a threefold criteria: the host star must be massive enough ($\gtrsim 0.32~M_\odot$) to place the HZ at orbital distances that prevent rapid planetary tidal synchronization, the HZ must be extended enough to sit outside the sub-Alfvénic stellar wind regime, and the planet's tidal dissipation must be sufficiently low to prevent accelerated rotational decay.

\section{Discussion\label{discussions}}

In our framework, we explicitly propagated the uncertainties related with the empirical stellar age estimations \citep{Engle2023}. Stellar age represents the most critical temporal constraint in our models, as it directly establishes the stage of tidal synchronization and the subsequent evolution of the planetary dynamo. While we use other empirical scaling laws in our framework, we adopted only their nominal value without the respective formal errors. Compounding the uncertainties across all empirical relations would artificially inflate the final error margins, obscuring the underlying trends.

As mentioned, the HZ of the low-mass stars lies completely in the sub-Alfvénic regime. In this environment, star-planet interactions are characterized by complex, direct magnetic connectivity, Alfvén wings, shocks, and enhanced Joule heating of the planetary atmosphere \citep[e.g.,][]{Laine2012,Cohen2014,Strugarek2015, Strugarek2018}. As these interactions are not considered, our calculated protection likelihoods likely represent a conservative upper bound. To capture the full severity of these environments 3D magnetohydrodynamic (MHD) simulations are required. Additionally, while several studies adopt scaling laws to model the evolution of the planetary magnetic field \citep{Griessmeier2005, 2007AsBio...7..185L, RodriguezMozos2019}, in reality, dynamos are driven by non-linear fluid dynamics governed by rotational shear, turbulent convection, density stratification and diffusivities \citep[e.g.][]{Ortiz2023, Kapyla2023, Hidalgo2024}. Nonetheless, given the current impossibility of characterizing exoplanetary cores, empirical scalings remain a robust framework for the estimation of planetary magnetic fields.

Throughout this work, the magnetopause is assumed to be in a steady state, and no transient phenomena like flares, and coronal mass ejections (CMEs) were considered. However, it is well known that M dwarfs are active stars and can exhibit a high rate of strong flares \citep{Loyd2018,Galleta2025}. The occurrence of CMEs in M dwarfs and their implications on habitability has historically been a topic of debate \citep{2007AsBio...7..167K,2007AsBio...7..185L,Wood2021}. However, only recently direct evidence has been obtained from radio observations
of an early M dwarf \citep{Callingham2025}. These detections imply that planets residing in the close-in habitable zones of M dwarfs are subject to frequent and intense CME impacts. Under such conditions, the transient dynamic pressures impacting the planet would likely be orders of magnitude higher than the steady ambient wind, driving severe transient compression of the magnetosphere and enhancing atmospheric erosion. Consequently, this is another factor that would only diminish the already critically low atmospheric protection likelihoods estimated in this study.

A strong initial planetary magnetic field could counterbalance the effects of synchronization. However, to achieve a protection likelihood of $100\%$, planets orbiting low-mass stars require initial field strengths significantly larger than those explored by our models, as shown in Figure~\ref{fig:req_B0p}. Finally, it should be noted that our framework assumes that the orbits have low eccentricities and are essentially circular. In the presence of a relevant non-zero eccentricity, 1:1 rotational synchronization is prevented, and the planet locks into a super-synchronous rotation state. This faster spin could maintain a stronger field in Model A and even prevent the dipolar collapse in Model B, until the orbits fully circularizes under the circularization timescale, which can range from millions to billions of years \citep{Barnes2017}.


\section{Conclusions \label{conclusions}}

We find that stellar tidal braking systematically reduces the planetary dipolar component, severely compromising magnetic shielding across the habitable zones of M dwarfs. Consequently, our results are significantly more pessimistic than previous studies modeling freely rotating Earth-like planets around M dwarfs \citep{Vidotto2013, RodriguezMozos2019}. At the HZ inner boundaries ($\ri^\mathrm{TL}$ and $\ri$), nearly our entire sample is tidally locked, which is in agreement with previous studies \citep{Griessmeier2005,Leconte2015, Barnes2017}. The rapid rotational synchronization ensures that planetary dynamos are either significantly attenuated (Model A) or driven into a multipolar regime due to critical local Rossby numbers (Model B) for both Modern (ME) and Early (EE) Earth scenarios, leaving atmospheres highly vulnerable to erosion. These negative effects are more prominent in low-mass stars ($0.093\text{--}0.39~M_\odot$), where the entire HZ is located deep within the sub-Alfvénic regime. This reinforces that sub-Alfvénic exposure is catastrophic for atmospheric retention, matching the findings of \cite{RodriguezMozos2019}.

At the outer HZ boundary ($\ro$), the less dissipative, rapidly rotating EE scenario exhibits a protection threshold at $M_\star \approx 0.32~M_\odot$. Below this stellar mass, the dipolar topology collapses and the protection likelihoods are essentially zero. For higher-mass hosts, EE planets successfully resist tidal synchronization and orbit safely outside the sub-Alfvénic stellar wind regime, maintaining robust $97\text{--}100\%$ protection likelihoods. In the ME scenario, the magnetic fields are severely affected in most of the sample due to the low rotation rates, while the super-Alfvénic conditions around high-mass hosts provide a non-zero protection likelihood, this shielding is typically weaker than in the EE case. Consequently, we show that a shielded atmospheric environment around M dwarfs requires a host massive enough to place the HZ at wider orbital distances, avoiding the effects of the sub-Alfvénic regime, and the internal tidal dissipation of the planet must be sufficiently low to prevent rapid rotational decay.

 We have assessed complex processes including the interplay of tidal locking with planetary dynamos and the implications for their interactions with stellar winds. It is of course clear that within an analytic framework, these effects can only be treated in a simplified way. Nonetheless they highlight the necessity of coupling tidal locking and the planetary dynamos to evaluate habitability. 3D MHD simulations should be employed in the future to investigate the non-linear interaction between the stellar winds and the complex, multipolar fields of tidally locked planets.

\begin{acknowledgements}
We thank the anonymous referee for providing useful and
constructive comments on the manuscript. JPH thanks Paulina Karczmarek for insights on plotting, clear data presentation and beyond. JPH acknowledges financial support from ANID/DOCTORADO BECAS CHILE 72240057. DRGS gratefully acknowledges support by the ANID BASAL project FB21003 and via the  Alexander von Humboldt - Foundation, Bonn, Germany.
\end{acknowledgements}

\bibliographystyle{aa}
\bibliography{astro.bib}

\begin{appendix}

\onecolumn

\section{Extra tables}

\renewcommand{\arraystretch}{1.1} 
\begin{table*}[h!]
\centering
\caption{Stellar parameters of the M dwarf sample from Section~\ref{real-examples}.}
\begin{tabular}{ccccccccccc}
\hline\hline
Name & SpT & $M_\star~[M_\odot]$ & $R_\star~[R_\odot]$ & $\log(L_\star/L_\odot)$  & $P_\mathrm{rot}~[\mathrm{days}]$  & Ro & $B_0~[\mathrm{G}]$ & $t_\star ~[\mathrm{Gyr}]$ & HZ [AU] & Ref\\ 
\hline
WX UMa &  M6.0   & 0.10 & 0.12 & -3.12 &  0.78 & 0.005 & 980 & $0.72^{+0.05}_{-0.05}$ & $(0.02) 0.03, 0.06$ & 2\\
GJ 1245B &  M5.5   & 0.12 & 0.14 & -2.83  & 0.71 & 0.005 & 136 & $0.72^{+0.05}_{-0.05}$ & $(0.03) 0.04, 0.08$ & 2 \\
GL Vir &  M5.0   & 0.14 & 0.16 & -2.64  & 0.49 & 0.004 & 86 & $0.71^{+0.05}_{-0.05}$ & $(0.04) 0.05, 0.10$ & 2 \\
V388 Cas &  M5.0   & 0.22 & 0.22 & -2.34  & 1.06 & 0.010 & 1613 & $0.73^{+0.05}_{-0.05}$ & $(0.05) 0.07, 0.14$ & 2 \\
EQ Peg B &  M4.5   & 0.25 & 0.25 & -2.15  & 0.40 & 0.004 & 450 & $0.71^{+0.05}_{-0.05}$ & $(0.07) 0.09, 0.17$ & 4 \\
V374 Peg &  M4.0   & 0.28 & 0.28 & -2.02  & 0.45 & 0.005 & 710 & $0.71^{+0.05}_{-0.05}$ & $(0.08) 0.10, 0.20$ & 3\\
EV Lac &  M3.5   & 0.32 & 0.30 & -1.92 &  4.38 & 0.057 & 530 & $0.23^{+0.01}_{-0.01}$ & $(0.09) 0.11, 0.22$ & 4\\
EQ Peg A &  M3.5   & 0.39 & 0.35 & -1.77 & 1.06 & 0.017 & 480 & $0.15^{+0.01}_{-0.01}$ & $(0.10) 0.14, 0.26$ & 4\\
AD Leo &  M3.5   & 0.42 & 0.38 & -1.69  & 2.24 & 0.039 & 250 & $0.17^{+0.01}_{-0.01}$ & $(0.11) 0.15, 0.29$ & 4\\
CE Boo &  M2.5   & 0.48 & 0.43 & -1.57  & 14.70 & 0.306 & 103 & $0.86^{+0.05}_{-0.05}$ & $(0.13) 0.17, 0.33$ & 1\\
OT Ser &  M1.5   & 0.55 & 0.49 & -1.40  & 3.40 & 0.086 & 130 & $0.15^{+0.01}_{-0.01}$ & $(0.15) 0.21, 0.40$ & 1\\
GJ 49 &  M1.5   & 0.57 & 0.51 & -1.33 & 18.60 & 0.494 & 27 & $1.29^{+0.19}_{-0.16}$ & $(0.16) 0.22, 0.43$ & 1\\
DS Leo &  M1.0   & 0.58 & 0.52 & -1.31  & 14.00 & 0.382 & 84 & $0.67^{+0.08}_{-0.07}$ & $(0.17) 0.23, 0.44$ & 1\\
DT Vir &  M0.5   & 0.59 & 0.53 & -1.28  & 2.85 & 0.080 & 147 & $0.14^{+0.01}_{-0.01}$ & $(0.17) 0.24, 0.45$ & 1\\
V1005 Ori &  M0.0   & 0.75 & 0.82 & -0.83  & 4.35 & 0.182 & 172 & $0.17^{+0.02}_{-0.01}$ & $(0.29) 0.40, 0.75$ & 1\\
\hline
\end{tabular} 
\tablefoot{From left to right: star name, spectral type, stellar mass $M_\star$, stellar radius $R_\star$, luminosity $L_\star$, rotation period $P_\mathrm{rot}$, Rossby number Ro, average surface magnetic field $B_0$, stellar age $t_\star$, habitable zone $(\ri^\mathrm{TL})\ri,\ro$. \\
\textbf{References.} 1, \cite{Donati2008a}; 2, \cite{Morin2010}; 3, \cite{2008MNRAS.384...77M}; 4, \cite{2008MNRAS.390..567M}.}
\label{table-stellar-real-cases}
\end{table*}

\renewcommand{\arraystretch}{1.1} 
   \begin{table*}[h]
   \centering
    \caption{Magnetospheres and magnetic fields of planets at the HZ.}
\begin{tabular}{cc|cc|cc}
\hline\hline
\multicolumn{2}{c}{M dwarfs} & \multicolumn{2}{c}{Model A} & \multicolumn{2}{c}{Model B}
\\
\hline
& &  \multicolumn{4}{c}{Modern-Earth at $(\ri^\mathrm{TL})\ri, \ro$} 
\\
Name & Regime & $r_\mathrm{M}/r_\mathrm{pl}$ & $B_\mathrm{p,0}^{(\omega)}$ [G] & $r_\mathrm{M}/r_\mathrm{pl}$ & $B_\mathrm{p,0}^{(\omega)}$ [G] \\ 
\hline
WX UMa & S & $(1.07 ) 1.12  , 1.23 $ &  $ (0.25 )  0.18 , 0.06 $  & $(0.79 ) 0.92  , 1.44 $ &  $ (0.10 )  0.10 , 0.10 $ \\
GJ 1245B & S & $(1.78 ) 1.71  , 1.64 $ &  $ (0.17 )  0.12 , 0.04 $  & $(1.48 ) 1.62  , 2.19 $ &  $ (0.10 )  0.10 , 0.10 $ \\
GL Vir & S & $(1.69 ) 1.70  , 1.63 $ &  $ (0.14 )  0.09 , 0.03 $  & $(1.52 ) 1.75  , 2.35 $ &  $ (0.10 )  0.10 , 0.10 $ \\
V388 Cas & S & $(0.80 ) 0.83  , 0.89 $ &  $ (0.10 )  0.07 , 0.02 $  & $(0.79 ) 0.94  , 1.42 $ &  $ (0.10 )  0.10 , 0.10 $ \\
EQ Peg B & S & $(1.03 ) 0.95  , 0.71 $ &  $ (0.08 )  0.05 , 0.02 $  & $(1.11 ) 1.18  , 1.22 $ &  $ (0.10 )  0.10 , 0.10 $ \\
V374 Peg & S & $(0.91 ) 0.87  , 0.66 $ &  $ (0.07 )  0.04 , 0.02 $  & $(1.03 ) 1.13  , 1.21 $ &  $ (0.10 )  0.10 , 0.10 $ \\
EV Lac & S & $(1.06 ) 1.09  , 1.08 $ &  $ (0.06 )  0.04 , 0.02 $  & $(1.24 ) 1.47  , 2.04 $ &  $ (0.10 )  0.10 , 0.10 $ \\
EQ Peg A & S & $(0.99 ) 0.95  , 1.04^{+0.04}_{-0.03}$ &  $ (0.05 )  0.03 , 0.04 $  & $(1.22 ) 1.35  , 1.47 $ &  $ (0.10 )  0.10 , 0.10 $ \\
AD Leo & S & $(1.14 ) 1.10  , 1.82^{+0.06}_{-0.06}$ &  $ (0.05 )  0.03 , 0.06^{+0.01}_{-0.01}$  & $(1.46 ) 1.62  , 2.16 $ &  $ (0.10 )  0.10 , 0.10 $ \\
CE Boo & NS & $(1.47 ) 1.46  , 1.37^{+0.01}_{-0.00}$ &  $ (0.04 )  0.03 , 0.01 $  & $(1.98 ) 2.26  , 2.94 $ &  $ (0.10 )  0.10 , 0.10 $ \\
OT Ser & S & $(1.15 ) 1.12  , 4.18^{+0.07}_{-0.08}$ &  $ (0.03 )  0.02 , 0.54^{+0.03}_{-0.03}$  & $(1.66 ) 1.87  , 2.38 $ &  $ (0.10 )  0.10 , 0.10 $ \\
GJ 49 & NS & $(1.72 ) 1.65  , 2.31^+0.31_{-0.27}$ &  $ (0.03 )  0.02 , 0.03^+0.01_{-0.01}$  & $(2.54 ) 2.83  , 3.55 $ &  $ (0.10 )  0.10 , 0.10 $ \\
DS Leo & NS & $(1.46 ) 1.43  , 3.89^{+0.25}_{-0.26}$ &  $ (0.03 )  0.02 , 0.18^{+0.04}_{-0.03}$  & $(2.19 ) 2.49  , 3.21 $ &  $ (0.10 )  0.10 , 0.10 $ \\
DT Vir & S & $(1.08 ) 1.06  , 4.63^{+0.04}_{-0.04}$ &  $ (0.03 )  0.02 , 0.73^{+0.02}_{-0.02}$  & $(1.63 ) 1.86  , 3.12^{+0.44}_{-0.36}$ &  $ (0.10 )  0.10 , 0.23^{+0.11}_{-0.07}$ \\
V1005 Ori & NS & $(0.86 ) 2.57^{+0.10}_{-0.11} , 5.30 $ &  $ (0.02 )  0.26^{+0.03}_{-0.03}, 0.97 $  & $(1.60 ) 1.88  , 5.35 $ &  $ (0.10 )  0.10 , 1.00 $ \\
\hline
& &  \multicolumn{4}{c}{Early-Earth at $(\ri^\mathrm{TL})\ri, \ro$} 
\\
Name & Regime & $r_\mathrm{M}/r_\mathrm{pl}$ & $B_\mathrm{p,0}^{(\omega)}$ [G] & $r_\mathrm{M}/r_\mathrm{pl}$ & $B_\mathrm{p,0}^{(\omega)}$ [G] \\ 
\hline
WX UMa & S & $(0.78 ) 0.81  , 0.89 $ &  $ (0.10 )  0.07 , 0.02 $  & $(0.70 ) 0.82  , 1.28 $ &  $ (0.07 )  0.07 , 0.07 $ \\
GJ 1245B & S & $(1.29 ) 1.24  , 1.19 $ &  $ (0.07 )  0.04 , 0.02 $  & $(1.32 ) 1.44  , 1.95 $ &  $ (0.07 )  0.07 , 0.07 $ \\
GL Vir & S & $(1.22 ) 1.23  , 1.18 $ &  $ (0.05 )  0.03 , 0.01 $  & $(1.35 ) 1.55  , 2.09 $ &  $ (0.07 )  0.07 , 0.07 $ \\
V388 Cas & S & $(0.58 ) 0.60  , 0.64 $ &  $ (0.04 )  0.03 , 0.01 $  & $(0.70 ) 0.84  , 1.26 $ &  $ (0.07 )  0.07 , 0.07 $ \\
EQ Peg B & S & $(0.75 ) 0.69  , 0.53^{+0.01}_{-0.01}$ &  $ (0.03 )  0.02 , 0.01 $  & $(0.98 ) 1.04  , 1.09 $ &  $ (0.07 )  0.07 , 0.07 $ \\
V374 Peg & S & $(0.66 ) 0.63  , 0.79^{+0.05}_{-0.05}$ &  $ (0.03 )  0.02 , 0.03^{+0.01}_{-0.00}$  & $(0.92 ) 1.00  , 1.07 $ &  $ (0.07 )  0.07 , 0.07 $ \\
EV Lac & S & $(0.77 ) 0.79  , 3.02^{+0.03}_{-0.03}$ &  $ (0.02 )  0.02 , 0.33^{+0.01}_{-0.01}$  & $(1.10 ) 1.30  , 3.81^{+0.05}_{-0.11}$ &  $ (0.07 )  0.07 , 0.65^{+0.03}_{-0.06}$ \\
EQ Peg A & S & $(0.72 ) 0.69  , 2.59^{+0.01}_{-0.01}$ &  $ (0.02 )  0.01 , 0.54^{+0.01}_{-0.01}$  & $(1.09 ) 1.20  , 2.82 $ &  $ (0.07 )  0.07 , 0.70 $ \\
AD Leo & S & $(0.83 ) 0.79  , 3.86^{+0.01}_{-0.01}$ &  $ (0.02 )  0.01 , 0.57^{+0.00}_{-0.01}$  & $(1.30 ) 1.44  , 4.13 $ &  $ (0.07 )  0.07 , 0.70 $ \\
CE Boo & NS & $(1.07 ) 1.05  , 4.64^{+0.05}_{-0.05}$ &  $ (0.02 )  0.01 , 0.39^{+0.01}_{-0.01}$  & $(1.75 ) 2.00  , 5.63 $ &  $ (0.07 )  0.07 , 0.70 $ \\
OT Ser & S & $(0.83 ) 1.81^{+0.10}_{-0.11} , 4.50^{+0.01}_{-0.01}$ &  $ (0.01 )  0.09^{+0.02}_{-0.01}, 0.67 $  & $(1.47 ) 1.66  , 4.56 $ &  $ (0.07 )  0.07 , 0.70 $ \\
GJ 49 & NS & $(1.24 ) 1.19  , 6.22^{+0.07}_{-0.08}$ &  $ (0.01 )  0.01 , 0.54^{+0.02}_{-0.02}$  & $(2.26 ) 2.52  , 6.78 $ &  $ (0.07 )  0.07 , 0.70 $ \\
DS Leo & NS & $(1.06 ) 1.12^{+0.07}_{-0.04} , 5.90^{+0.03}_{-0.03}$ &  $ (0.01 )  0.01 , 0.62^{+0.01}_{-0.01}$  & $(1.94 ) 2.21  , 6.13 $ &  $ (0.07 )  0.07 , 0.70 $ \\
DT Vir & S & $(0.81^{+0.02}_{-0.01}) 2.53^{+0.07}_{-0.08} , 4.52 $ &  $ (0.01 )  0.25^{+0.02}_{-0.02}, 0.69 $  & $(1.45 ) 1.69^{+0.40}_{-0.04} , 4.55 $ &  $ (0.07 )  0.08^{+0.07}_{-0.01}, 0.70 $ \\
V1005 Ori & NS & $(2.52^{+0.04}_{-0.05}) 3.48^{+0.01}_{-0.01} , 4.75 $ &  $ (0.39^{+0.02}_{-0.02})  0.64^{+0.01}_{-0.01}, 0.70 $  & $(3.07 ) 3.59  , 4.75 $ &  $ (0.70 )  0.70 , 0.70 $ \\
\hline
\end{tabular} 
    \tablefoot{Same as Table~\ref{table-planet} but for the stars in Section~\ref{real-examples}.}
    \label{table-real-cases-planet}
    \end{table*}

   \begin{table*}[h]
   \centering
    \caption{Atmospheric protection likelihoods.}
    \begin{tabular}{c|cc|cc}
    \hline\hline
     & \multicolumn{2}{c}{Modern Earth} & \multicolumn{2}{c}{Early Earth}
    \\
    M dwarfs  & $P_\mathrm{A}(\alpha_0)[(\ri^\mathrm{
    TL}) \ri, \ro]$ & $P_\mathrm{B}(\alpha_0)[(\ri^\mathrm{
    TL}) \ri, \ro]$  & $P_\mathrm{A}(\alpha_0)[(\ri^\mathrm
    TL) \ri, \ro]$ & $P_\mathrm{B}(\alpha_0)[(\ri^\mathrm{
    TL}) \ri, \ro]$ \\ 
    \hline
M6.5 S & ($0.00 $)$0.00 $, $0.00 $ &  ($0.00 $) $0.00 $, $0.00 $ &  ($ 0.00$)$0.00 $, $0.00 $ &  ($ 0.00$) $0.00 $, $0.00 $ \\
M4.5 S & ($0.00 $)$0.00 $, $0.00 $ &  ($0.00 $) $0.00 $, $0.02 $ &  ($ 0.00$)$ 0.00$, $ 0.00$ &  ($0.00 $) $0.00 $, $0.00 $ \\
M3.5 S& ($0.00 $)$0.00 $, $0.00 $ &  ($0.00 $) $0.00 $, $0.03 $ &  ($ 0.00$)$ 0.00$, $0.09^{+0.02}_{-0.02}$ &  ($0.00 $) $0.00 $, $0.01 $ \\
M1.5 S& ($0.00 $)$0.00 $, $0.91^{+0.02}_{-0.02}$ &  ($0.00 $) $0.01 $, $0.09 $ &  ($ 0.00$)$0.00 $, $0.98 $ &  ($0.00 $) $0.00 $, $0.98 $ \\
M0.0 S& ($0.00 $)$0.00 $, $1.00 $ &  ($0.00 $) $0.01 $, $1.00 $ &  ($0.00 $)$0.50^{+0.02}_{-0.03}$, $0.99 $ &  ($0.00 $) $0.76 $, $0.99 $ \\
M6.5 NS & ($0.05 $)$0.07 $, $0.12 $ &  ($0.00 $) $0.01 $, $0.36 $ &  ($0.00 $)$0.00 $, $0.00 $ &  ($0.00 $) $0.00 $, $0.19 $ \\
M4.5 NS & ($0.01 $)$0.01 $, $0.02 $ &  ($0.01 $) $0.05 $, $0.53 $ &  ($0.00 $)$0.00 $, $0.00 $ &  ($0.00 $) $0.01 $, $0.32 $ \\
M3.5 NS & ($0.00 $)$0.00 $, $0.00 $ &  ($0.02 $) $0.08 $, $0.57 $ &  ($0.00 $)$0.00 $, $0.00 $ &  ($0.00 $) $0.02 $, $0.36 $ \\
M1.5 NS & ($0.00 $)$0.00 $, $0.00 $ &  ($0.16 $) $0.30 $, $0.70 $ &  ($0.00 $)$0.00 $, $1.00 $ &  ($0.06 $) $0.14 $, $1.00 $ \\
M0.0 NS & ($0.00 $)$0.00 $, $1.00 $ &  ($0.17 $) $0.32 $, $0.70 $ &  ($0.00 $)$0.00 $, $1.00 $ &  ($0.07 $) $0.15 $, $1.00 $ \\
    \hline
WX UMa & ($0.00 $)$0.00 $, $0.00 $ &  ($ 0.00$) $ 0.00$, $0.00 $ &  ($ 0.00$)$ 0.00$, $ 0.00$ &  ($ 0.00$) $ 0.00$, $0.00 $ \\
GJ 1245B & ($0.00 $)$0.00 $, $0.00 $ &  ($0.00 $) $0.00 $, $0.04 $ &  ($0.00 $)$0.00 $, $0.00 $ &  ($0.00 $) $0.00 $, $0.01 $ \\
GL Vir & ($0.00 $)$0.00 $, $0.00 $ &  ($0.00 $) $0.00 $, $0.07 $ &  ($0.00 $)$0.00 $, $0.00 $ &  ($0.00 $) $0.00 $, $0.02 $ \\
V388 Cas & ($ 0.00$)$ 0.00$, $ 0.00$ &  ($ 0.00$) $ 0.00$, $0.00 $ &  ($ 0.00$)$ 0.00$, $ 0.00$ &  ($ 0.00$) $ 0.00$, $0.00 $ \\
EQ Peg B & ($0.00 $)$ 0.00$, $ 0.00$ &  ($0.00 $) $0.00 $, $0.00 $ &  ($ 0.00$)$ 0.00$, $ 0.00$ &  ($ 0.00$) $0.00 $, $0.00 $ \\
V374 Peg & ($ 0.00$)$ 0.00$, $ 0.00$ &  ($0.00 $) $0.00 $, $0.00 $ &  ($ 0.00$)$ 0.00$, $ 0.00$ &  ($ 0.00$) $0.00 $, $0.00 $ \\
EV Lac & ($0.00 $)$0.00 $, $0.00 $ &  ($0.00 $) $0.00 $, $0.01 $ &  ($ 0.00$)$ 0.00$, $0.38^{+0.02}_{-0.02}$ &  ($0.00 $) $0.00 $, $0.79^{+0.02}_{-0.05}$ \\
EQ Peg A & ($ 0.00$)$ 0.00$, $0.00 $ &  ($0.00 $) $0.00 $, $0.00 $ &  ($ 0.00$)$ 0.00$, $0.16 $ &  ($0.00 $) $0.00 $, $0.27 $ \\
AD Leo & ($0.00 $)$0.00 $, $0.00 $ &  ($0.00 $) $0.00 $, $0.03 $ &  ($ 0.00$)$ 0.00$, $0.81 $ &  ($0.00 $) $0.00 $, $0.89 $ \\
CE Boo & ($0.00 $)$0.00 $, $0.00 $ &  ($0.01 $) $0.05 $, $0.33 $ &  ($0.00 $)$0.00 $, $0.98^{+0.00}_{-0.01}$ &  ($0.00 $) $0.01 $, $1.00 $ \\
OT Ser & ($0.00 $)$0.00 $, $0.91^{+0.02}_{-0.02}$ &  ($0.00 $) $0.00 $, $0.08 $ &  ($ 0.00$)$0.00 $, $0.97 $ &  ($0.00 $) $0.00 $, $0.98 $ \\
GJ 49 & ($0.00 $)$0.00 $, $0.06^{+0.11}_{-0.05}$ &  ($0.14 $) $0.27 $, $0.67 $ &  ($0.00 $)$0.00 $, $1.00 $ &  ($0.05 $) $0.13 $, $1.00 $ \\
DS Leo & ($0.00 $)$0.00 $, $0.82^{+0.08}_{-0.11}$ &  ($0.03 $) $0.12 $, $0.49 $ &  ($0.00 $)$0.00 $, $1.00 $ &  ($0.01 $) $0.04 $, $1.00 $ \\
DT Vir & ($0.00 $)$0.00 $, $0.98 $ &  ($0.00 $) $0.00 $, $0.44^{+0.24}_{-0.20}$ &  ($ 0.00$)$0.13^{+0.03}_{-0.03}$, $0.97 $ &  ($0.00 $) $0.00^{+0.02}_{-0.00}$, $0.98 $ \\
V1005 Ori & ($ 0.00$)$0.15^{+0.05}_{-0.04}$, $1.00 $ &  ($0.00 $) $0.00 $, $1.00 $ &  ($0.13^{+0.02}_{-0.02}$)$0.64^{+0.00}_{-0.01}$, $0.99 $ &  ($0.40 $) $0.69 $, $0.99 $ \\
    
    \hline
    \end{tabular} 
    \tablefoot{Atmospheric protection likelihood $P(\alpha_0)$ of planets orbiting the HZ: [$(\ri^\mathrm{TL})\ri,\ro$] of the theoretical and the real samples. The underscore “A” represents Model A, and “B” model B.}
    \label{table-probabilities}
    \end{table*}

\end{appendix}

\end{document}